\documentclass{ieeeaccess}
\usepackage{cite}
\usepackage{amsmath,amssymb,amsfonts}
\usepackage{algorithmic}
\usepackage{graphicx}
\usepackage{textcomp}
\def\BibTeX{{\rm B\kern-.05em{\sc i\kern-.025em b}\kern-.08em
    T\kern-.1667em\lower.7ex\hbox{E}\kern-.125emX}}

\newcommand{\yes}{\text{\Large$\bullet$}}%
\newcommand{\no}{\text{\Large$\circ$}}%

\newcommand{\server}{S}
\newcommand{\doctor}{C}
\newcommand{\reader}{R}

\newcommand{\simdprotocol}{IMDfence\xspace}

\newcommand{\sTL}{T\xspace}

\newcommand{\sConf}{SR1\xspace}
\newcommand{\sAvailability}{SR2\xspace}
\newcommand{\sNR}{SR3\xspace}
\newcommand{\sEA}{SR4\xspace}
\newcommand{\sMulti}{SR5\xspace}
\newcommand{\sAC}{SR6\xspace}
\newcommand{\sAuth}{SR7\xspace}
\newcommand{\sFlexibility}{SR8\xspace}
\newcommand{\sBedside}{SR9\xspace}
\newcommand{\valid}{R_{valid}}
\newcommand{\stolen}{R_{stolen}}
\newcommand{\hacked}{R_{hacked}}
\newcommand{\forged}{R_{forged}}
\newcommand{\leguserValid}{S1\xspace}
\newcommand{\anyuserStolen}{S2\xspace}
\newcommand{\leguserForged}{S3\xspace}
\newcommand{\maluserHacked}{S4\xspace}
\newcommand{\maluserForged}{S5\xspace}
\newcommand{\attackerValid}{S6\xspace}
\newcommand{\attackerForged}{S7\xspace}

\usepackage{xspace}
\usepackage{threeparttable}
\usepackage{cleveref}
\usepackage{booktabs} 

\newcommand\blfootnote[1]{%
	\begingroup
	\renewcommand\thefootnote{}\footnote{#1}%
	\addtocounter{footnote}{-1}%
	\endgroup
}

\begin{document}
\history{Date of publication xxxx 00, 0000, date of current version xxxx 00, 0000.}
\doi{10.1109/ACCESS.2020.3015686}

\title{\simdprotocol: Architecting a Secure Protocol for Implantable Medical Devices}
\author{\uppercase{Muhammad Ali Siddiqi}\authorrefmark{1}, \uppercase{Christian Doerr}\authorrefmark{2},
\uppercase{and Christos Strydis}\authorrefmark{1},
\IEEEmembership{Senior Member, IEEE}}
\address[1]{Department of Neuroscience, Erasmus Medical Center, Rotterdam, The Netherlands (e-mail: m.siddiqi@erasmusmc.nl, c.strydis@erasmusmc.nl)}
\address[2]{Cyber Threat Intelligence Lab, Hasso Plattner Institute, University of Potsdam, Germany (e-mail: christian.doerr@hpi.de)}
\tfootnote{This work has been supported by the EU-funded project SDK4ED (Grant Agreement No. 780572).}

\markboth
{Siddiqi \headeretal: \simdprotocol: Architecting a Secure Protocol for Implantable Medical Devices}
{Siddiqi \headeretal: \simdprotocol: Architecting a Secure Protocol for Implantable Medical Devices}

\corresp{Corresponding author: Muhammad Ali Siddiqi (e-mail: m.siddiqi@erasmusmc.nl).}

\begin{abstract}
    Over the past decade, focus on the security and privacy aspects of implantable medical devices (IMDs) has intensified, driven by the multitude of cybersecurity vulnerabilities found in various existing devices.
    However, due to their strict computational, energy and physical constraints, conventional security protocols are not directly applicable to IMDs. Custom-tailored schemes have been proposed instead which, however, fail to cover the full spectrum of security features that modern IMDs and their ecosystems so critically require.
    In this paper we propose \simdprotocol, a security protocol for IMD ecosystems that provides a comprehensive yet practical security portfolio, which includes availability, non-repudiation, access control, entity authentication, remote monitoring and system scalability.
    The protocol also allows emergency access that results in the graceful degradation of offered services without compromising security and patient safety.
    The performance of the security protocol as well as its feasibility and impact on modern IMDs are extensively analyzed and evaluated.
    We find that \simdprotocol achieves the above security requirements at a mere less than 7\% increase in total IMD energy consumption, and less than 14 ms and 9 kB increase in system delay and memory footprint, respectively.
\end{abstract}

\begin{keywords}
Authentication protocol, battery-depletion attack, battery DoS, denial-of-service attack, IMD, implantable medical device, non-repudiation, smart card, zero-power defense
\end{keywords}

\titlepgskip=-15pt

\maketitle

\blfootnote{This article has been accepted for publication in a future issue of this journal, but has not been fully edited. Content may change prior to final publication. Citation information: DOI 10.1109/ACCESS.2020.3015686, IEEE Access. This work is licensed under a Creative Commons Attribution 4.0 License. For more information, see https://creativecommons.org/licenses/by/4.0/.}

\section{Introduction}
\label{sec:introduction}

Modern implantable medical devices (IMDs), such as cardiac pacemakers and defibrillators, neurostimulators, and more, are equipped with wireless connectivity in order to aid in treatment-related reconfiguration, patient-health monitoring, device testing etc.~\cite{siddiqi2018attack,seepers2016implantable}.
However, wireless links have made IMDs susceptible to various attacks by malicious entities.

Earlier-generation IMDs had little or no security provisions whatsoever, as confirmed by numerous ethical-hacking incidents over the past decade~\cite{halperin2008pacemakers,marin2016security,marin2018securing}.
The research community has responded with a wealth of new schemes and, eventually, top IMD manufacturers now claim to have rectified the security weaknesses over the past few years~\cite{medtronic2019bulletin,fda2019abbott}.

However, due to the constraints imposed by an IMD's scant computational, storage and energy resources, most proposed schemes in research have refrained from taking proven security approaches.
Moreover, since these schemes have been specifically tailored for IMDs, they have missed the big picture and resulted in limited coverage of the security properties essential to a modern IMD. Specifically, most focus has been drawn on confidentiality, integrity, authentication and emergency access (e.g.,~\cite{bu2019bulwark,belkhouja2019biometric,chi2018safe,belkhouja2018symmetric} etc.), while non-repudiation, remote monitoring and system scalability have been left unaddressed for the most part.
Besides being difficult to tackle, prior seminal work has not identified or stressed the importance of these additional requirements.

In this paper, we debunk the myth that advanced security is impossible in modern IMDs. To this end, we collect both well-studied and overlooked security requirements, impose strict design constraints, and propose \simdprotocol, a novel security protocol for IMD ecosystems.
This works contributes:

\begin{itemize}
	\item A comprehensive security protocol for a modern IMD ecosystem, \simdprotocol, which addresses crucial, yet previously ignored requirements, i.e., non-repudiation, remote monitoring and system scalability.
	\item A realistic solution for accessing the IMD during emergencies without compromising security or patient safety.
	\item A rigorous evaluation of \simdprotocol paying special attention to the protection against battery denial-of-service (DoS) attacks.
\end{itemize}

The rest of the paper is organized as follows:
We enumerate modern IMD-system requirements in Section~\ref{sec:requirements}, and then discuss existing systems and related works in Sections~\ref{sec:existing-systems} and~\ref{sec:related-work}, respectively.
Section~\ref{sec:imd-protocol} details our proposed security protocol.
We evaluate \simdprotocol in Section~\ref{sec:evaluation} and provide concluding remarks in Section~\ref{sec:conclusion}.

\section{IMD-security requirements}
\label{sec:requirements}

In this section, we collect and present the necessary security and related functional requirements that should be satisfied in modern IMD systems.
These requirements form the basis of the IMD-specific security protocol, to be detailed in Section~\ref{sec:imd-protocol}.

In order to evaluate the IMD-system security, we consider an implant that is capable of communicating wirelessly with a reader/programmer\footnote{The term \textit{reader} will be used for any device that is able to directly communicate with the implant.}.
We assume an attacker whose aim could be to either (1) modify or sabotage IMD operation in order to prevent patient treatment, (2) manipulate patient-related data, or (3) steal patient data.
Furthermore, we assume that the attacker has \textit{full control} of the wireless channel between the reader and IMD. This means that he/she can eavesdrop, modify, insert, block or replay messages between these two entities at will.
As a result, the IMD-security system has to satisfy certain security requirements (SRs):

\subsection{Basic security services (\sConf \& \sAvailability)}

As in other domains, the IMD-security system should provide the fundamental security services: \textit{Confidentiality}, \textit{Integrity} and \textit{Availability}. The first two services (\sConf) are usually addressed through the use of lightweight block-ciphers and message-authentication codes (MAC)~\cite{strydis2013system}.
More specifically, the commands sent from the reader to the IMD and the associated responses (e.g., data logs) should be treated as confidential and it should be ensured that such data is not modified in transit.

Availability ensures that the IMD is always available for patient treatment whenever required (\sAvailability).
This implies that the device should be protected against Denial-of-Service (DoS) attacks.
One of the highest-likelihood and lowest-cost attacks is the battery-depletion attack (or battery DoS attack), as indicated in the IMD-specific threat-modeling analysis in~\cite{siddiqi2018attack} and practically demonstrated in~\cite{halperin2008pacemakers,marin2016security}.

\subsection{Non-repudiation (\sNR)}

Non-repudiation ensures that the sender of a message is not able to deny (or \textit{repudiate}) its creation.
Since there is always a possibility of malpractices, medical mistakes or insider attacks, we require non-repudiation to aid in computer forensics in case a patient experiences medical issues as a direct consequence of such actions.
This security service ensures that a physician, paramedic or nurse is not able to deny his/her involvement in such scenarios.
Non-repudiation has not been given due consideration by the research community when it comes to IMD systems.
One of the reasons is that \textit{true} non-repudiation can only be achieved through the use of public-key (or asymmetric) cryptography for computing digital signatures~\cite{van2016emv}, which has traditionally been considered to be too resource-costly for IMDs~\cite{strydis2013system,siddiqi2019imd}.
Another, very important, reason is that past generations of IMDs could only be accessed by one person, i.e., the physician. Nowadays, the IMDs can be accessed by multiple people, including the patients themselves~\cite{medtronic2017pacemaker,stjude2016confirm,proclaim2017proclaim}. Hence, there is a need to introduce \textit{user accountability}.

Most of the existing IMD-security works have looked into strict reader-IMD communication (without the involvement of a trusted third party).
Even if we assume that the resource-constrained IMD is able to support public-key computations, this reader-IMD configuration makes it impossible for the IMD to effectively use public-key cryptography since it cannot keep track of the validity of the reader certificates (due to lack of Internet connectivity).
What is more, these devices do not have sufficient memory to store the required certificates~\cite{marin2018security}.
For instance, the IMD must store all possible reader certificates if we want to support access during travels or when the patient is visiting abroad.
Hence, a scheme is required that employs \textit{additional} architectural components (as will be discussed in Section~\ref{sec:imd-protocol}) to solve these issues.

Another complication is the legal aspect. Since non-repudiation is there to provide evidence, it should be incorporated based on the assumption that such evidence will be scrutinized by a hostile legal expert~\cite{anderson1994liability}.
One main limitation of cryptography-based non-repudiation is that there is no formally-verifiable link between the device that signs the digital signature and its user.
For example, the user, i.e., the private-key owner, can falsely claim that the signature has been generated by a malware program without his/her consent, or that the private key has been stolen.
There is no technical mechanism that can determine whether such a claim is false~\cite{roe2010cryptography}.
The IMD security protocol should address this limitation, which we term as the \textit{Non-repudiation gap}. 

\subsection{Emergency Access (\sEA)}

Patient safety always outweighs device security. Hence, during emergencies the security protocol should not hinder or delay paramedic access to the IMD~\cite{rostami2013heart,seepers2016implantable}.
Although it seems reasonable to drop security altogether in such situations, this can be a problem if, while in a normal mode, an adversary fools the IMD into entering the emergency-access mode.
The security protocol must be capable of allowing the IMD to accurately \textit{classify} whether a communication attempt is an emergency or a normal access.
This ensures that the adversary is unable to trigger and exploit the emergency-access mode.
Furthermore, since there is a high likelihood of the patient losing control of his/her actions in emergencies, the emergency-access mode should be independent of patient participation.

\subsection{Multi-manufacturer environment (\sMulti)}

Past works on emergency access have ignored the fact that, in emergencies, it is unlikely for the paramedic to know the IMD make and model beforehand. Moreover, it is not possible to preemptively stock all the readers from all the manufacturers in the ambulance.
Hence, to achieve \textit{true} emergency access, the IMD-security system should be \textit{manufacturer-independent}, i.e., all manufacturers need to agree on a unified standard for secure reader-IMD communication.
This way, an ambulance can use one generic reader regardless of the IMD manufacturer and type.
It follows that an emergency-access scheme should be adoptable by all IMD types. E.g., an emergency-access solution that requires an IMD measuring the cardiac signal~\cite{rostami2013heart}, can be easily incorporated in pacemakers, but it will require significant modifications in neurostimulators.

As things stand, true emergency access does not exist in commercial IMDs.
As long as this remains acceptable to the medical community, \sMulti can be relaxed. This is further discussed in Section~\ref{sec:without-standardization}.

\subsection{Access control (\sAC)}
\label{sec:access-control}

The access privileges of the reader should be differentiated based on the type of user.
For example, nurses, patients or patient relatives may only be allowed to read status data from the implant, whereas a physician and a paramedic may further be allowed to modify the implant configuration for therapy updates, suspend or resume its operation. Similarly, a technician may be allowed to modify the implant firmware in addition to tasks of the above user roles.

\subsection{User and reader-IMD Authentication (\sAuth)}

In order to aid in non-repudiation and access control, the IMD system should be able to identify the physician, nurse, paramedic etc. who is using the reader to communicate with the implant.
Similarly, the reader should also be able to authenticate the IMD in order to prevent spoofing attacks on the reader.
Hence, there is a requirement for performing \textit{mutual} authentication instead of just authenticating the reader unilaterally~\cite{strydis2013system}.
Furthermore, said authentication is required to be \textit{strong}, i.e., it should imply both \textit{message} and \textit{entity} authentication, and guarantee message \textit{freshness}, or in other words \textit{replay protection}.

\subsection{Flexibility and Scalability (\sFlexibility)}

The IMD should not be limited to communicating with only a fixed amount of readers since this severely limits portability, e.g., during emergencies when a paramedic reader is used, or when there is a need for treatment at some hospital during travels. Hence, there should not be any pre-shared secrets between the reader and IMD.

\subsection{Bedside-reader operation for remote monitoring (\sBedside)}

Some of the modern IMD systems also include a bedside reader, which enables remote monitoring~\cite{merlin2015faq}. It establishes communication with the IMD when the patient is asleep and sends treatment status to a back-end server via an Internet connection.
However, this additional connection represents an increase in the attack surface, which imposes additional security requirements.
We predict that the use of such readers will become more widespread over time due to their time- and cost-saving features. Hence, this phenomenon should proactively be considered when designing secure IMD systems.

\section{Existing systems}
\label{sec:existing-systems}

IMD manufacturers have typically relied on ``security through obscurity''; they choose to hide the communication-protocol specifications in order to enhance security.
This is not a recommended practice, and as a consequence of using this approach, we have seen several successful blackbox-hacking attempts over the past few years~\cite{halperin2008pacemakers,marin2016security}.

Some of the latest commercial IMDs, including neurostimulators~\cite{proclaim2017proclaim}, insertable cardiac monitors~\cite{stjude2016confirm} and even pacemakers~\cite{medtronic2017pacemaker} offer a Bluetooth Low Energy (BLE) connection between the patient smart-phone and the implant. The initial pairing between these devices is based on the BLE standard in addition to proprietary protocols~\cite{proclaim2017proclaim}. However, they do not disclose the \textit{association models} used in these pairings, which makes these devices vulnerable to attacks due to the reasons mentioned above.
In most of the cardiac devices, in the absence of an IMD-programmer, a magnet can be used to disable therapy or to switch to a default behavior~\cite{stjude2017icd}.
This mode, however, can be easily exploited by adversaries through the use of a strong magnet when in close proximity to the patient (e.g., in public transport).

\section{Related work}
\label{sec:related-work}

\begin{table*}[!t]
	\centering
	\begin{threeparttable}
		\caption{Overview of related works}
		\label{table:related-works}
		\footnotesize
		\begin{tabular}{lcccccccccc}
			\toprule
			\
			& \cite{bu2019bulwark} & \cite{chi2018safe} & \cite{belkhouja2018symmetric} & \cite{wazid2018novel} & \cite{mao2018trusted} & \cite{rathore2018multi}& \cite{fu2019pok}& \cite{ellouze2018powerless} & \cite{camara2020access} & \cite{park2014security}\\
			\midrule
			Confidentiality \& Integrity (\sConf)	&   \yes  &  \yes &  \yes &  \yes &  \yes &  \yes &  \yes &	 \yes &  \yes & \yes \\
			Availability (\sAvailability)  			&   \no	  &  \no  &  \no  &  \no  &  \no  &  \no  &  \no  &	 \yes &	 \no  & \no  \\
			Non-repudiation (\sNR)  				&   \no	  &  \no  &  \no  &  \no  &  \no  &  \no  &  \no  &	 \no  &	 \no  & \yes \\
			Emergency Access (\sEA)		  			&   \yes  &  \yes &  \no  &  \no  &  \no  &  \yes &  \yes &	 \yes &  \yes & \yes \\
			Multi-manufacturer support (\sMulti)	&   \no	  &  \no  &  \no  &  \no  &  \no  &  \no  &  \no  &	 \no  &	 \no  & \no  \\
			Access Control (\sAC)		  			&   \yes  &  \yes &  \no  &  \yes &  \yes &  \yes &  \yes &	 \no  &	 \no  & \yes \\
			Authentication (\sAuth)  				&   \yes  &  \yes &  \yes &  \yes &  \yes &  \yes &  \yes &	 \yes &  \yes & \yes \\
			Flexibility \& Scalability (\sFlexibility)&   \no &  \yes &  \no  &  \no  &  \no  &  \no  &  \no  &	 \yes &  \yes & \no  \\
			Beside-reader operation (\sBedside) 	&   \no	  &  \no  &  \no  &  \no  &  \no  &  \no  &  \no  &	 \no  &	 \no  & \no  \\
			\bottomrule
		\end{tabular}
		\begin{tablenotes}
			\footnotesize
			\item \yes: Satisfies requirement, \no: Does not satisfy requirement
		\end{tablenotes}
	\end{threeparttable}
\end{table*}

From the perspective of the research community, we observe a steep rise in the number of works proposed over the last few years~\cite{rushanan2014sok}.
For data confidentiality, integrity and message authentication, the use of lightweight primitives has been proposed.
Early works focused on basic security protocols based on symmetric ciphers, which rely on a common pre-shared key between the reader and the IMD~\cite{strydis2013system}.
However, such approaches are not scalable in terms of adding new readers that can access the implant.
They also do not allow paramedic access during emergencies.
Therefore, most of the existing works deal with emergency access, in addition to entity authentication and key exchange.
For entity authentication, these works rely on a \textit{touch-to-access} policy, which ensures that only the entities that can physically touch the patient for a prolonged period of time are allowed access to the implant~\cite{rostami2013heart,siddiqi2018attack}.
In other words, it is infeasible for an attacker to get in close proximity to the patient, and even if that is the case, the patient can detect this and reject physical contact.
Also, the attacker would then have far easier methods to harm the patient than via accessing the implant, e.g., by physically attacking the patient.
These works can be broadly categorized as follows~\cite{seepers2016implantable}:

\textbf{Biometric-based:} These approaches (such as~\cite{rostami2013heart,seepers2016secure}) rely on both the reader and IMD to measure a physiological signal from different parts of the patient's body. The devices are paired based on the similarity of these measurements.

\textbf{Proxy-based:} These works propose to use an additional device in the possession of the patient, such as a smart phone, watch, etc~\cite{pournaghshband2012securing,sorber2012amulet}. The device is paired with the IMD and is used to authenticate the reader that is trying to communicate with the implant. In case of emergency, the device can be physically distanced from the patient in order to grant the reader unsecured access to the IMD.

\textbf{Distance-based:} These works (e.g.,~\cite{rasmussen2009proximity,kim2015vibration}) employ weak or out-of-band (OOB) signals for reader-IMD communication. These can either involve direct transfer of a session key, which would be hard for an attacker to eavesdrop, or they can require the devices to mutually prove proximity to one another.

\textbf{Token-based:} This is the simplest approach, which relies on the patients having the IMD-access key or password with them, which is stored e.g., on a bracelet.
During an emergency, a paramedic can access the IMD using this token.

We now present a brief overview of the latest works from literature that were specifically tailored for IMDs.

Bu et al. propose a low-energy IMD-security scheme called Bulwark~\cite{bu2019bulwark}, which, in addition to satisfying \sConf, also allows IMD access in emergencies (\sEA).
This emergency access scheme is based on Shamir's \textit{secret sharing}, which relies on the users (including the paramedics) to register with the manufacturer of the specific IMD in advance in order to retrieve the access key in case of an emergency. As evident, such a requirement inhibits IMD access in case the patient is out of town (\sFlexibility).

Chi et al.~\cite{chi2018safe} propose a protocol that relies on the patient's smartphone for the reader access. However, requiring the patient to be in possession of this additional device (i.e., the smartphone) all the time, including during emergencies, puts a significant burden on the patient.

Belkhouja et al.~\cite{belkhouja2018symmetric} propose a symmetric crypto system in which they use a \textit{Chaotic key generator} that is employed by both the reader and IMD to generate the symmetric key. However, in order for this key generator to work, both entities are required to have similar pre-installed initial conditions/values. Hence, this scheme cannot function in an emergency scenario, or when the patient is traveling, since the IMD and the reader will not be sharing the same initial conditions.

Wazid et al.~\cite{wazid2018novel} and Mao et al.~\cite{mao2018trusted} propose three-factor protocols, which rely on passwords, smart cards, and biometrics.
Their protocols rely on a reader-registration phase before the IMD deployment in the field.
This inhibits \sEA and \sFlexibility since it is unlikely for the paramedic/doctor to possess a pre-registered reader during an emergency or when the patient is visiting abroad.
Rathore et al.~\cite{rathore2018multi} propose a scheme in which the identifiers of each user (including the patient) are derived from their cardiac signals and are stored in the implant.
Hence, it requires a user-registration phase similar to the above protocols.
However, their scheme allows emergency access since the paramedic can measure patient's cardiac signal, which is compared by the IMD against the stored identifier in order to grant access.
The three-factor protocol from Fu et al.~\cite{fu2019pok} also provides emergency access. However, the patient is required to always be in possession of a personal smart card so that the paramedic is able to use it during an emergency.

A few works~\cite{halperin2008pacemakers,strydis2013system,ellouze2018powerless} have also focused on the IMD availability (\sAvailability). In these works, RF energy harvesting is employed to protect the IMD against battery-depletion attacks.
In addition, quite a few authentication and emergency-access schemes have been proposed recently that rely on static biometrics (such as fingerprints)~\cite{zheng2018finger}, dynamic biometrics (such as cardiac signals)~\cite{ellouze2018powerless,camara2020access} and combination of both~\cite{belkhouja2019biometric}. The interested reader can refer to~\cite{wu2017access,zheng2016ideas,altawy2016security,camara2015security,rushanan2014sok} to get an overview of prior works in this area.

Overall, the above works address only parts of the IMD security requirements (\sConf, \sAvailability, \sEA, \sAC, \sAuth and \sFlexibility), which is also summarized in Table~\ref{table:related-works}. For instance, non-repudiation is not considered and the emergency-access schemes do not take into account the (current) multi-manufacturer environment, as discussed in Section~\ref{sec:requirements}.
To the best of our knowledge, there is no protocol that provides all the services highlighted in Section~\ref{sec:requirements}.

The work from literature that came closest to fulfilling the above requirements was proposed by Park~\cite{park2014security}. It establishes a session key between the IMD and a \textit{personalized} reader based on shared secrets between these entities and a trusted third party (hospital server).
The use of public-key crypto in the personalized reader and the server facilitates non-repudiation.
However, the work lacks a few additional pieces in order to properly close the non-repudiation gap (as will be discussed in Section~\ref{sec:addressing-nr-gap}).
The protocol addresses access control by first allowing only read access to the implant via the server. Based on the result of the read-out data, the server provides write keys to the reader-IMD pair which allows the user to change IMD settings.
The personalization process involves the physician inserting a personal smart card into the reader.
However, since it resembles a single-factor authentication for the user (i.e., through the use of a smart card without PIN), any person in possession of a valid (stolen) card can access the implant by getting hold of a reader.
The server maintains a list of primary-care physicians authorized to access each registered implant. If the physician is a member of this list, then a read-key is granted to the physician. We believe that maintaining such a user list is not scalable, it inhibits flexibility, and hence, should not be employed. As an example, such a scheme will not work in case the patient requires some treatment at a hospital abroad.
Besides, the proposed emergency-access scheme uses a bracelet that has a secret key.
However, such token-based security schemes are single points of failure (e.g., in case the token is stolen or the contents are disclosed). Also, it requires the patient to wear the bracelet at all times, which is inconvenient. Moreover, in the emergency scenario, the scheme drops access control and non-repudiation.
Lastly, this work excludes battery DoS from its adversarial model, and it does not consider bedside-reader operation. 

\section{\simdprotocol: Security Protocol for IMD Ecosystems}
\label{sec:imd-protocol}


\Figure[!t]()[trim={2cm 4.5cm 2cm 4.5cm},clip,scale=0.38]{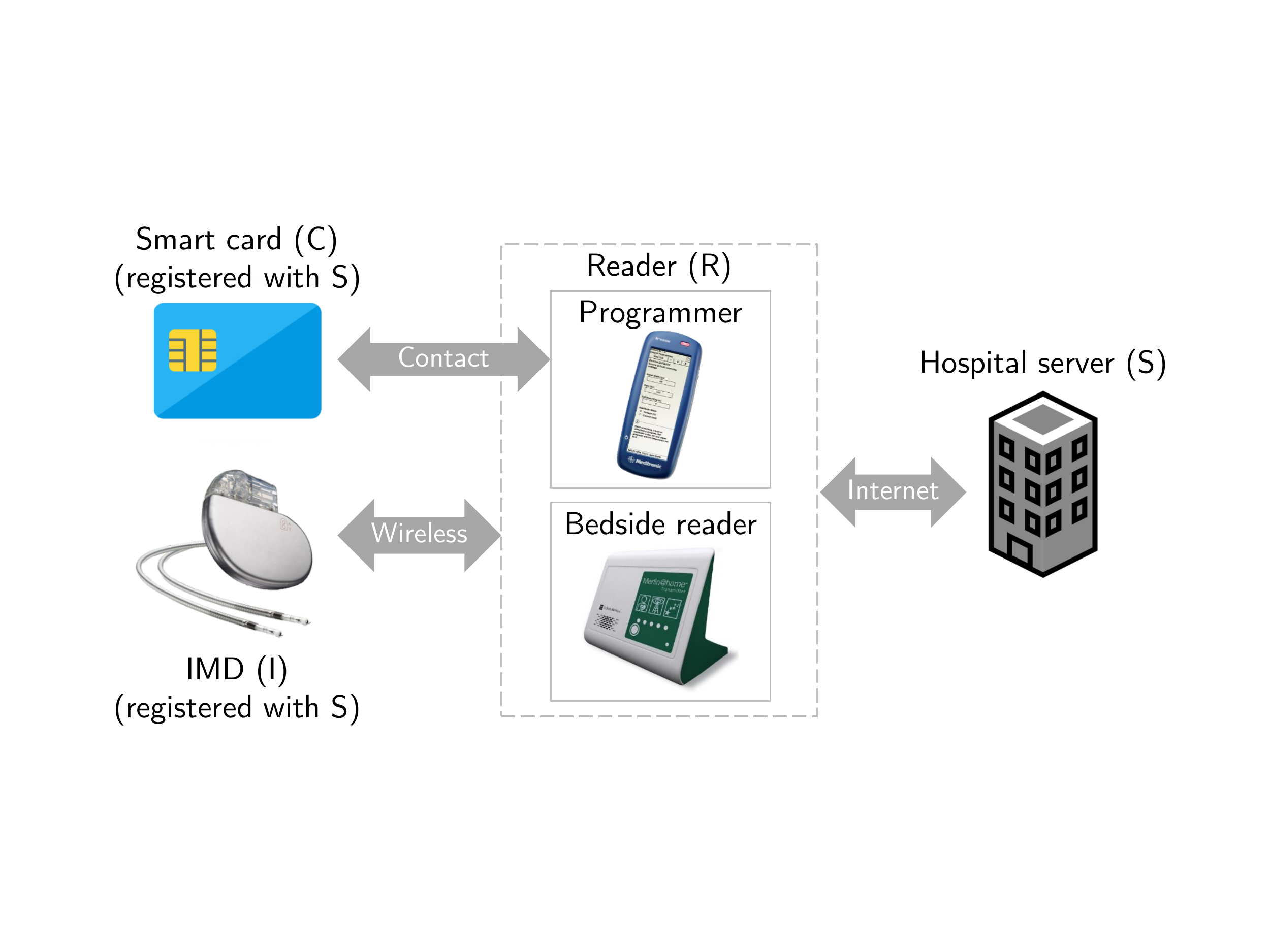}
{Proposed IMD ecosystem\label{fig:imd-emv-system}}

\begin{table}
	\centering
	\begin{threeparttable}
	\caption{Table of Notations}
	\label{table:table_of_notations}
	\small
	\begin{tabular}{p{45pt}p{170pt}}
		\toprule
		\
		Notation 	& Definition	\\
		\midrule
		$ID_A$	& ID of entity $A$ \\
		$N_A$	& Nonce generated by $A$ \\
		$K_{pubA}$/$K_{prA}$ & Public/private key pair of $A$ \\
		$K_{AB}$ & Pre-shared symmetric key between $A$ and $B$\\
		$K'_{AB}$ & Short-term symmetric key between $A$ and $B$\\
		$K_{A}$ & A secret only known to $A$\\
		$P_A$	& Privilege information of user/card $A$ \\
		$CMD$ & Configuration command\\
		$ANS$ & Answer to $CMD$ \\
		$k$ & Difficulty of solving a client puzzle\\
		$x<i>$ & $i^{th}$ bit of a bitstring $x$\\
		$x<i:j>$ & Bit sequence $x<i>,...,x<j>$\\
		$t$ & Time stamp\\
		$\sTL$ & Lifetime of reader-card authentication\\
		$\{\}_{K_{AB}}$ & Authenticated encryption\tnote{*} using $K_{AB}$ \\
		$\mathsf{MAC}_{K_{AB}}()$ & MAC operation using key $K_{AB}$ \\
		$\mathsf{sig}_{K_{prA}}()$	& Signature (of a hashed message) using $K_{prA}$ \\
		$Cert_A$ & Certificate consisting of $ID_A$, $P_A$, $K_{pubA}$ and $\mathsf{sig}_{K_{prCA}}(ID_A, P_A, K_{pubA})$ \\
		\bottomrule
	\end{tabular}
	\begin{tablenotes}
	\footnotesize
	\item [*]Such as Encrypt-then-MAC (EtM). Separate keys should be used for the encryption and MAC operations to prevent certain attacks and to ease key management~\cite{roe2010cryptography}. However, these keys are not differentiated here for simplicity.
	\end{tablenotes}
\end{threeparttable}
\end{table}

The absence of a complete security solution for IMD systems has led us to propose \simdprotocol, a novel secure-communication protocol that satisfies the extensive and strict requirements enumerated in Section~\ref{sec:requirements}. As will be shown, \simdprotocol addresses the complete IMD ecosystem.

\subsection{Configuration and assumptions}
\label{sec:protocol_assumptions}

The \simdprotocol configuration includes a smart card
($\doctor$) for the user ($U$) trying to access the IMD (e.g., a physician), and a trusted third party (TTP), i.e., a hospital server ($\server$), in addition to the implant ($I$) and the reader ($R$); see Fig.~\ref{fig:imd-emv-system}.
The list of notations used in this paper is summarized in Table~\ref{table:table_of_notations}.
The extra components, $\doctor$ and $\server$, are employed to facilitate \textit{\textbf{non-repudiation (\sNR), access control (\sAC)}} and \textit{\textbf{user authentication (\sAuth)}}, as identified in Section~\ref{sec:requirements}.
Each personal smart card, which is inserted in $R$, supports public-key cryptography. Its private key, which is unique to each card/user, enables digital-signature computation, thus providing non-repudiation.
Since $R$ and $\doctor$ are untrusted with respect to each other, a TTP ($\server$) is required to mutually authenticate the two entities.
Non-repudiation can technically also be provided through the use of a \textit{personal} reader that supports public-key computations in order to get rid of $\doctor$ and $\server$.
However, such a solution would be highly impractical and expensive since it would require all the doctors and nurses to be in possession of their personal readers at all times.
Moreover, the use of $\server$ also enables access control and facilitates \textit{\textbf{bedside-reader operation (\sBedside)}}.
Every user requires their own $\doctor$ and should know the associated PIN (two-factor authentication).
Since patients are only allowed \textit{read-only} access (as discussed in Section~\ref{sec:access-control}), losing or misplacing their $\doctor$ will not inhibit any future treatment. 
To avoid additional attack vectors, we propose to not support the use of \textit{contactless} smart cards and magnetic-strip cards.

\subsubsection{Interfaces}
\label{sec:protocol_interfaces}

For tackling \textit{\textbf{flexibility and scalability (\sFlexibility)}}, there is no pre-shared key between $R$ $\leftrightarrow$ $I$, $R$ $\leftrightarrow$ $\doctor$, $\server$ $\leftrightarrow$ $R$, and $\doctor$ $\leftrightarrow$ $I$.
The only pre-shared symmetric keys that exist are between $\server$ $\leftrightarrow$ $I$ ($K_{\server{I}}$) and $\server$ $\leftrightarrow$ $\doctor$ ($K_{\server{\doctor}}$).
A unique $K_{\server{I}}$ is installed in the implant at the time of manufacturing, which is then shared with the server of the hospital where the implantation surgery is going to take place.
During this IMD-registration process, the implant is also assigned a unique and random identifier $ID_I$, which is stored in the implant.
Likewise, $K_{\server{\doctor}}$ is installed in the smart card and is shared with the hospital where the card user is registered.
Moreover, $\server$, $I$ and $\doctor$ can only talk to $R$ directly and only indirectly with each other\footnote{The routing details of the messages communicated via the reader have been omitted for brevity.}.

The secure communication between $\server$ $\leftrightarrow$ $R$ is made possible by employing public-key-based key exchange in which the public/private key pairs of these entities are used.
This configuration helps in making $R$ independent of the need to pre-share keys with the hospital, which aids in scalability.
As a result, a patient can use his/her personal reader from any location, and/or buy a new reader from the manufacturer without the need of registering it first at the hospital.

In our proposed configuration, each smart card also has its own public/private key pair.
Technically, $R$ has the capability of maintaining a comprehensive certificate-revocation list (CRL) of smart cards due to frequent Internet connectivity.
Hence, it is able to verify smart-card certificates.
On the other hand, due to the limited on-board memory and less-frequent Internet connectivity, $\doctor$ can only maintain a small CRL that does not change frequently.
Hence, $\doctor$ can not verify the authenticity of the multitude of reader certificates.
As a result, public-key-based key exchange cannot be used to establish a session key between $R$ $\leftrightarrow$ $\doctor$.
However, it will be shown in Section~\ref{sec:regular_mode} that the session key between $R$ $\leftrightarrow$ $\doctor$ will be established using $\server$ as a TTP.
The same will be done for establishing a session key between $R$ $\leftrightarrow$ $I$.
Lastly, no session key is required between $\doctor$ $\leftrightarrow$ $I$.

\subsubsection{Centralization and Public-key infrastructure}
\label{sec:centralization}

The public keys of $\server$, $R$ and $\doctor$ are signed by a trusted certification authority (CA) belonging to the manufacturer.
The smart-card certificates, in addition, also include the user privileges.

We consider the precise implementation details of public-key infrastructure (PKI) and certificate revocation outside the scope of this paper.
In case of a smart card, certificate revocation would be needed when a card is stolen, a user leaves, or he/she changes roles (e.g., from nurse to paramedic).
For a reader, certificate revocation would be required in case $R$ is stolen or deemed as out-of-service.
The server is given the responsibility to verify the certificates of $R$ and $\doctor$ and hence, it is assumed that it maintains an up-to-date CRL.

\subsubsection{Modes of operation}

We propose two modes of operating in \simdprotocol, one for regular (online) operation and the other in the absence of an active Internet connection (offline), e.g., during \textit{\textbf{emergencies (\sEA)}}; see Fig.~\ref{fig:flow-chart}.
Online mode offers the full security- and functional-requirement portfolio highlighted in Section~\ref{sec:requirements}, whereas offline mode results in the graceful degradation of offered services without compromising security and patient safety.
Since $\server$ is not available in offline mode, $R$ and $I$ will be required to undergo an out-of-band (OOB) pairing phase in order to securely exchange a short-term session key.
These modes and the constituent phases will be elaborated in the following sections.


\Figure[!t]()[trim={2.8cm 5.5cm 2.8cm 5.5cm},clip,scale=0.42]{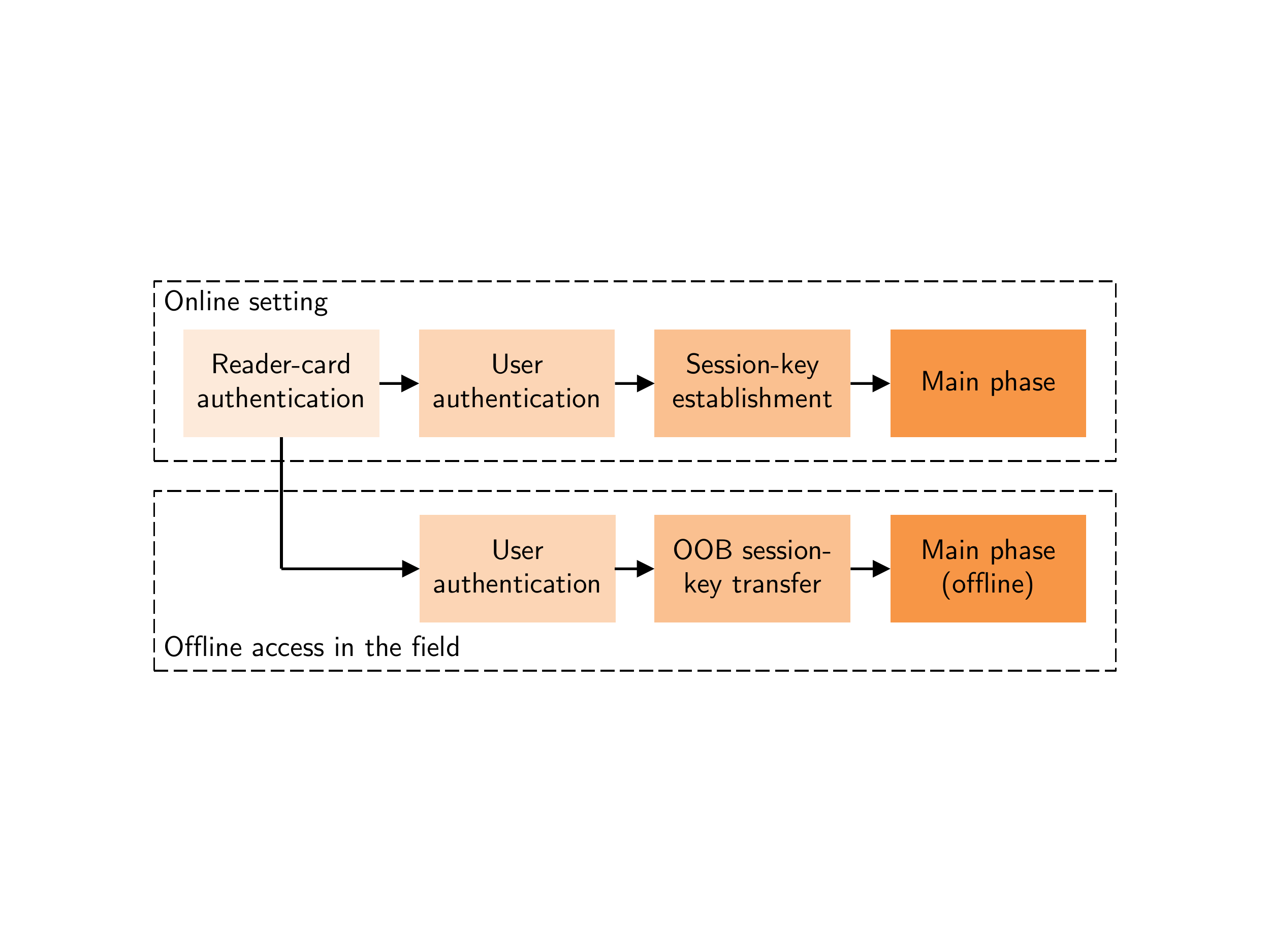}
{\simdprotocol flow under online and offline scenarios\label{fig:flow-chart}}

\subsection{Threat model}
\label{sec:threat_model}

As discussed in Section~\ref{sec:requirements}, we assume an attacker $A$ that has full control of the wireless channel between $R$ and $I$.
$R$ is assumed to be untrustworthy by $I$, $\doctor$ and $\server$, and vice versa.
Moreover, we assume that if $A$ steals a personal smart card or a valid reader, then the user or hospital staff should notify the hospital server so that it is blacklisted.
Additionally, we assume that $A$ can hack the reader to read out or modify data at the interface of the inserted smart card.
However, $A$ does not have access to the keys stored in $R$ and $\doctor$.
This implies that protection against side-channel attacks is considered outside the scope of this work since such attacks are typically addressed through specialized countermeasures.
Moreover, due to the assumption that $\server$ is notified of a lost/stolen device, $A$ has a limited time window to perform such attacks after stealing a device.
We also assume that the hospital personal do not have access to the keys stored in the server since such attacks can be prevented by employing standard practices, such as hardware security modules (HSM) etc.

\subsection{Regular (online) mode}
\label{sec:regular_mode}

The regular mode of \simdprotocol is shown in ~\Cref{fig:initial_kerberos,fig:user-auth,fig:kerberos,fig:main-phase}.
It starts with the \textit{$R$ $\leftrightarrow$ $\doctor$ mutual authentication} phase after the physician (or any other user) inserts their smart card into the reader.

\subsubsection{$R$ $\leftrightarrow$ $\doctor$ mutual authentication}


\Figure[!t]()[trim={0 0.5cm 0 1.3cm},clip,scale=0.61]{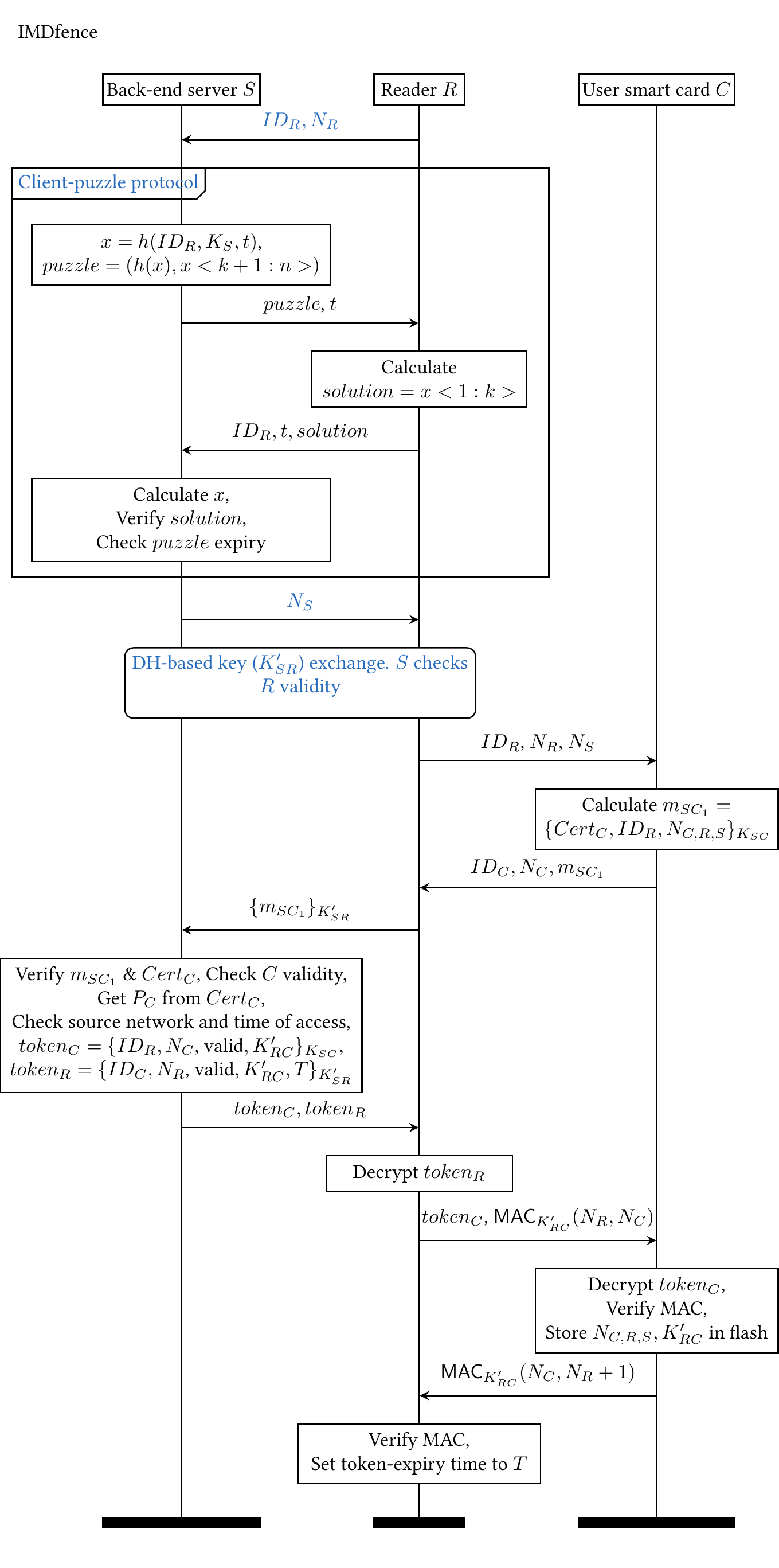}
{Reader-card authentication. Steps that are common with bedside-reader mode are marked in blue.\label{fig:initial_kerberos}\vspace{-0.4cm}}

In this phase, $R$ first tries to establish a secure connection with $S$ by sending its identifier and a nonce (which is a freshly generated number that is used only once).
In order to deter distributed-denial-of-service (DDoS) attacks against $S$ (to ensure server \textit{\textbf{availability (\sAvailability)}}), a basic client-puzzle protocol (CPP) is employed~\cite{juels1999client}.
CPP is a proof-of-work system in which any client (or in this case a reader) that wants to access the server (during high load) is required to correctly solve a cryptographic puzzle.
For a single client the costs of solving this puzzle are negligible.
However, in order to launch a successful DDoS by initiating a large number of simultaneous connections, it would be computationally infeasible for the attacker to solve a multitude of such puzzles.

$\server$ initiates CPP \textit{if} it senses a DDoS attack or it is dealing with an abnormally high number of simultaneous connections.
It first calculates $x$, which is the $n$-bit hash of $ID_R$, the current time stamp $t$ and its long-term secret $K_{\server}$.
It then computes a second hash ($h(x)$). $\server$ sends $h(x)$ and $x$ \textit{excluding} the first $k$ bits of $x$, along with the $t$.
$R$ computes the solution, i.e., the missing $k$ bits of $x$, and sends it along with $ID_R$ and the received time stamp.
$k$ represents the difficulty of solving the puzzle.
$\server$ calculates $x$ again and verifies that the solution indeed corresponds to the missing bits.
It also verifies, with the help of $t$, that the puzzle has not expired.
$\server$ is protected against memory exhaustion since it is not required to store any data for the verification of the puzzle solution.
In case these checks are successful, $\server$ sends its nonce to $R$.

$R$ then performs a Diffie--Hellman (DH)-based handshake with $\server$ in which a session key is established between them based on their public/private key pairs (see Fig.~\ref{fig:initial_kerberos}).
During this handshake, both verify each other's certificates and, additionally, $\server$ checks if $R$ is valid (i.e., it is not reported as stolen or out-of-service).

In order to achieve authentication between $R$ and $\doctor$, $R$ then initiates a five-pass, mutual-authentication protocol borrowed from the ISO/IEC 9798-2 standard~\cite{iso-9798} with $\server$ acting as a TTP (see Fig.~\ref{fig:initial_kerberos}).
$R$ and $\doctor$ ensure message freshness by exchanging their nonces in the first messages between them, and then verifying the existence of these nonces in the subsequent messages.
$R$ generates its nonce and sends it along with its identifier and $N_{\server}$ to $\doctor$.
$\doctor$ responds by generating $N_{\doctor}$ and sending a cryptogram ($m_{\server{\doctor}_1}$) that includes authenticated encryption of its certificate, $ID_R$ and nonces, along with $ID_{\doctor}$ and $N_{\doctor}$ in plaintext.
This cryptogram is calculated using $K_{\server{\doctor}}$ since it is intended for the server.
$R$ stores $ID_{\doctor}$ and $N_{\doctor}$, and forwards the cryptogram to the server, which establishes that it originated from $\doctor$ and that it is also tied to $R$.
The server then verifies $Cert_{\doctor}$ and checks the validity of $\doctor$, in case it has been reported stolen or has expired.
It then determines the required privileges ($P_{\doctor}$) for the particular user (e.g., physician, paramedic, nurse etc) from $Cert_{\doctor}$.
It also calculates tokens for both these entities using the respective symmetric keys.
These tokens include the nonces and identifiers of $R$ and $\doctor$ and a fresh symmetric key $K'_{R{\doctor}}$.
Additionally, $token_R$ also contains $\sTL$ (reader-card-authentication lifetime).
Based on these tokens, $R$ and $\doctor$ can ascertain each other's trustworthiness.

$R$ decrypts $token_R$, retrieves $K'_{R{\doctor}}$, calculates the MAC of the nonces, and forwards it along with $token_{\doctor}$ to $\doctor$.
The smart card similarly decrypts $token_{\doctor}$ and verifies the received MAC using $K'_{R{\doctor}}$.
It stores the nonces and $K'_{R{\doctor}}$ in its internal flash memory\footnote{There can be a time gap between this and the next stage (in offline mode). Since smart cards can only be powered by $R$, the above data has to be stored in the non-volatile (flash) memory so that $\doctor$ can be taken out of $R$ during this period.} so that it can verify and create messages in the subsequent stages.
$\doctor$ then sends a MAC that is calculated over $N_R$ and $N_\doctor$ (including an addition by $1$ to protect against replay of the previous message).
$R$ verifies the received MAC using $K'_{R{\doctor}}$.
At this point, both $R$ and $\doctor$ have mutually authenticated each other.

$R$ then sets its internal real-time clock to $\sTL$ and starts it to track the period over which the subsequent phases can execute without the need of reader-card authentication.
Since it is possible that $R$ is not connected to the Internet \textit{during} its operation (e.g., in emergencies), this scheme enforces that $R$, by design, shall only be usable for a certain duration until it has first established an Internet connection. This makes sure that $R$ receives critical firmware updates in time, if there are any. The selection and configuration of $\sTL$ will be discussed in Section~\ref{sec:t_and_nac}.

\subsubsection{User authentication}


\Figure[!t]()[trim={0 0.5cm 0 1.3cm},clip,scale=0.61]{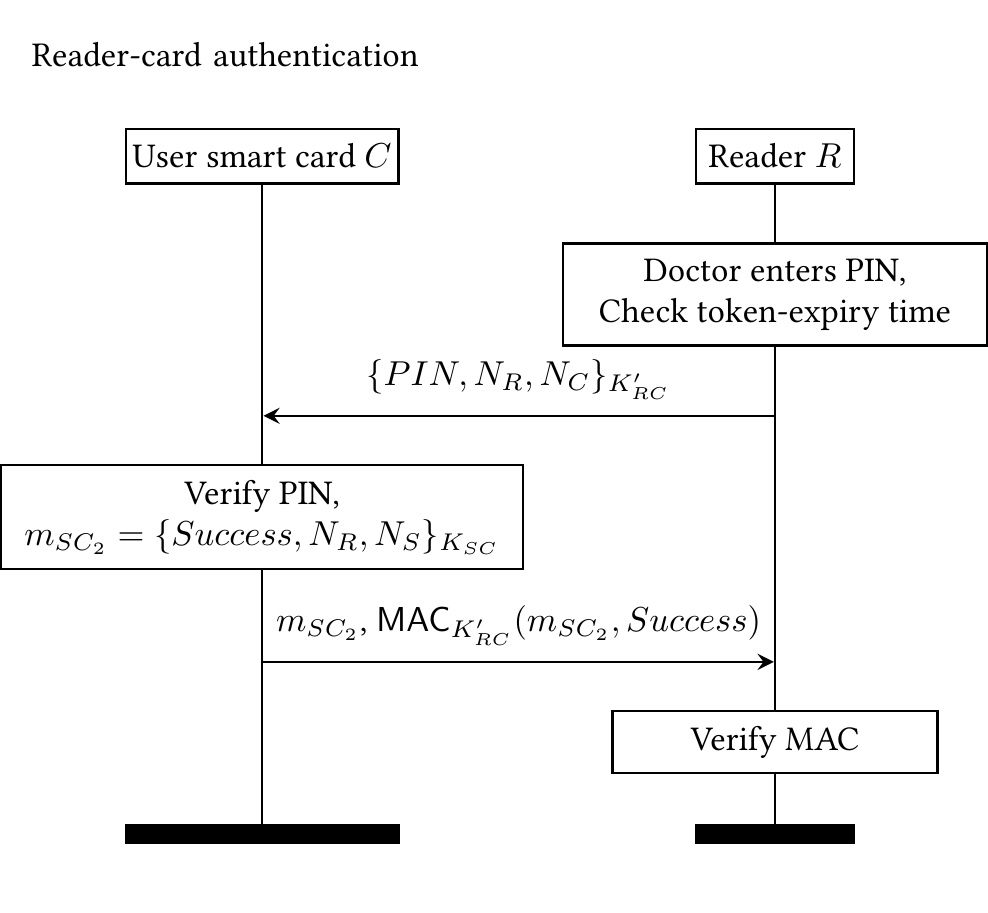}
{User authentication at the reader\label{fig:user-auth}}

This phase is shown in Fig.~\ref{fig:user-auth} and its objective is to authenticate the card holder.
The physician enters his/her PIN using a keypad on the reader.
$R$ then checks its internal real-time clock to verify the validity of its token.
$R$ encrypts the PIN and the nonces (in order to prevent replays) using ${K'_{R\doctor}}$. $\doctor$ decrypts the message using the same key, verifies the PIN by comparing it with the stored one and sends back a cryptogram intended for the server, which is encrypted with ${K_{\server\doctor}}$. It contains the confirmation of success in addition to the nonces.

\subsubsection{Session-key ($K'_{RI}$) establishment}
\label{sec:session-key-establishment}


\Figure[!t]()[trim={0 0.5cm 0 1.3cm},clip,scale=0.61]{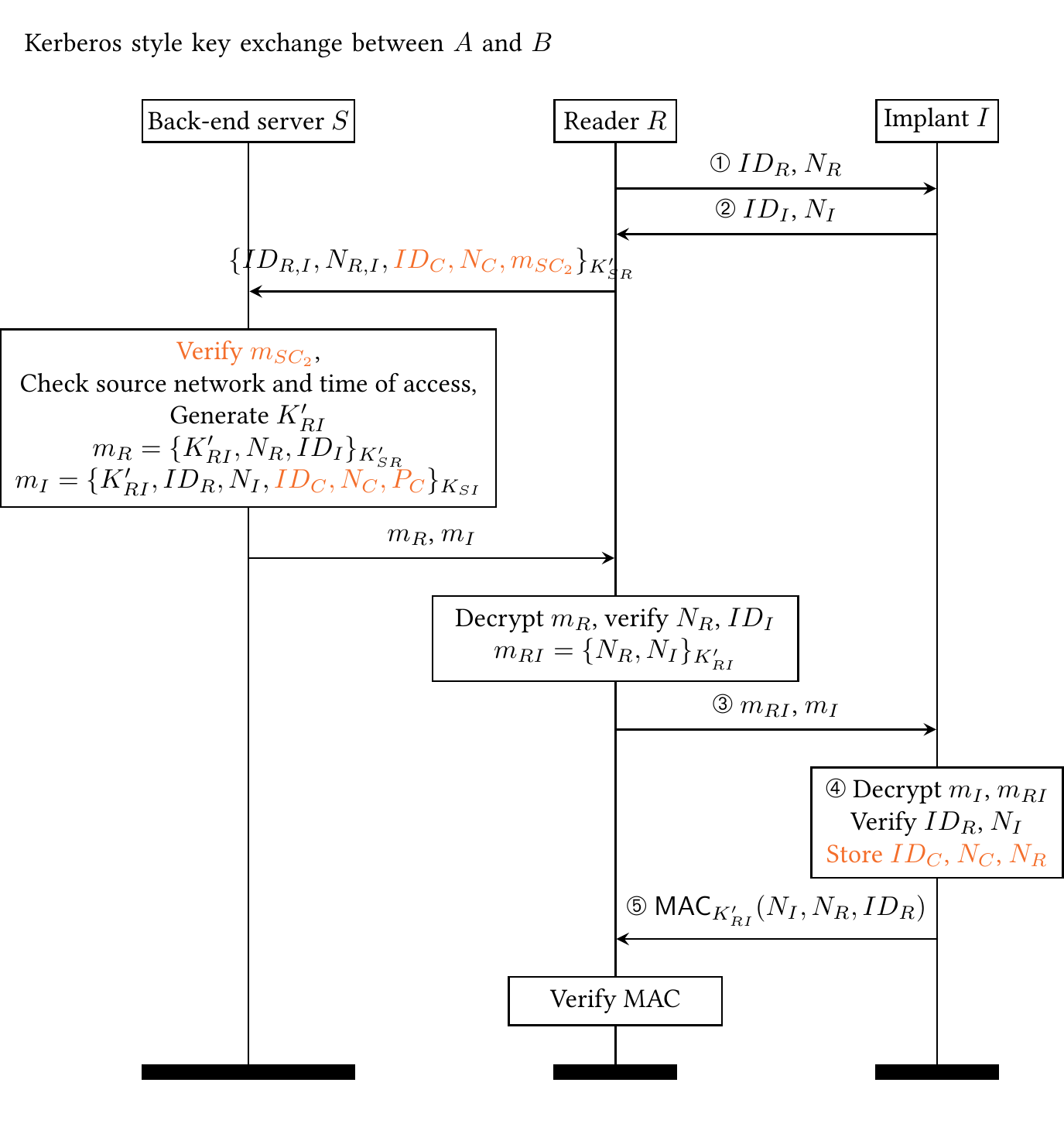}
{Session-key establishment between $R$ and $I$ via $\server$. Operations that are not relevant to bedside-reader mode are marked in orange.\label{fig:kerberos}}

$R$ then initiates a TTP-based key established protocol with $\server$ and $I$ in order to acquire a symmetric session key $K'_{RI}$ for providing \textit{\textbf{confidentiality and integrity (\sConf)}}, as shown in Fig.~\ref{fig:kerberos}.
$R$ first exchanges the nonces and identifiers with $I$ and then sends the nonces and identifiers of all parties to $\server$ along with $m_{\server\doctor_2}$.
$\server$ first verifies $m_{\server\doctor_2}$.
It then generates $K'_{RI}$, encrypts it in two independent messages $m_R$ and $m_I$ intended for $R$ and $I$ respectively, and then sends these to $R$.
$R$ decrypts $m_R$ and verifies its contents.
It then encrypts $N_R$ and $N_I$ using $K'_{RI}$ (to form $m_{RI}$) and then sends it along with $m_I$ to $I$.
$I$ first retrieves $K'_{RI}$ by decrypting $m_I$, and then decrypts $m_{RI}$ to verify that $R$ has the knowledge of $K'_{RI}$ and that the nonces are valid.
$I$ finally creates a MAC using the new session key for $R$ to validate.
At the end of this protocol, both $R$ and $I$ are mutually authenticated (\sAuth) and have arrived at a \textit{fresh} session key in addition to performing key confirmation.
Similar to the reader-card authentication stage, this phase is also based on the five-pass protocol from ISO/IEC 9798-2 since it involves a TTP.

To protect against battery-DoS attacks (which impact \textit{\textbf{availability (\sAvailability)}}), steps 1 to 4 of session-key establishment should be as lightweight as possible so that the IMD is able to execute it using harvested RF energy.
This will be further discussed in Section~\ref{sec:availability}.

\subsubsection{Main phase}
\label{sec:main-phase}


\Figure[!t]()[trim={0 0.5cm 0 1.3cm},clip,scale=0.61]{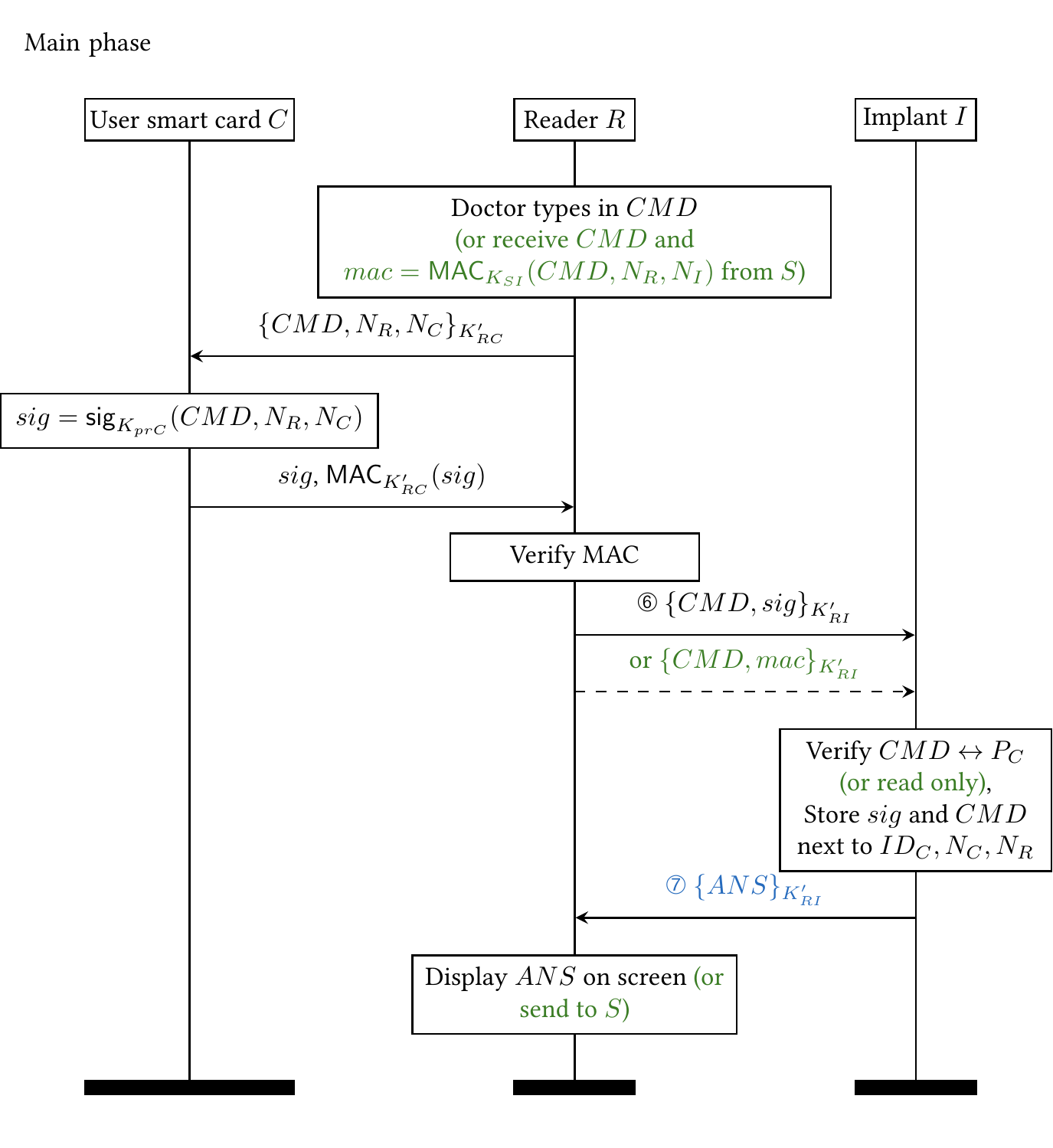}
{Main phase. Steps that are common with bedside-reader mode are marked in blue. Operations that are unique to bedside-reader mode are marked in green.\label{fig:main-phase}}

After session-key establishment, $R$ allows the user to enter a command on the reader interface (see Fig.~\ref{fig:main-phase}). The command is encrypted along with the nonces (to prevent replay attacks) using $K'_{R\doctor}$ and is sent to $\doctor$. The card decrypts the command, digitally signs the message using $K_{pr\doctor}$ (to form $sig$) and sends it to $R$.
$R$ re-encrypts the command using $K'_{RI}$ and sends it to the implant along with $sig$.

$I$ decrypts the command and verifies if it corresponds to the privileges information received in $m_I$ during the previous phase, hence ensuring \textit{access control}.
$sig$ and $CMD$ are stored by the IMD next to $ID_{\doctor}$, $N_{\doctor}$ and $N_R$, which were stored during session-key establishment. This is required to ensure non-repudiation since $sig$ was signed using a personal private key. For example, in the case of a medical mistake (e.g., an incorrect command) that led to patient death, the physician will not be able to deny his/her involvement since this signature can always be retrieved from the IMD and subsequently verified using the associated data. It follows that signature storage is not required for \textit{read-only} commands.
Since the implant trusts the reader at this point, there is no need for $I$ to verify the signature since the associated MAC has already been verified by $R$. This relieves $I$ of the need to employ public-key cryptography and to track user certificates.
After processing the command, the implant responds with an \textit{answer} message encrypted with $K'_{RI}$.
$R$ displays it on its screen for the convenience of the user.
The session keys expire after a \textit{finish} command and its associated response, or after a period $\sTL$.

\subsubsection{Addressing the non-repudiation gap}
\label{sec:addressing-nr-gap}

As discussed in Section~\ref{sec:requirements}, the use of a signature alone is not sufficient to address the legal aspects of non-repudiation.
In order to bridge the non-repudiation gap, one option could be to enforce that the user protects $\doctor$ and the associated PIN, or immediately reports in case it is lost.
However, due to the possibility of human error in general, this is too much of a legal responsibility for the user.

A realistic way of bridging this gap is by introducing additional checks in the implementation of reader-card-authentication and session-key-establishment phases (see Figures~\ref{fig:initial_kerberos} and~\ref{fig:kerberos}, respectively).
The server can ensure that the implant \textit{write} access (determined from $P_{\doctor}$) is requested from within the hospital network \textit{and} during the working hours of the user.
On the other hand, the server can allow \textit{read-only} accesses from external networks, e.g., in case the access is made by the patient or their bedside reader.
The user just has to ensure that $R$ is issued from a certified repository, \textit{and} that $R$ should only be connected to a trusted Ethernet/Wi-Fi network (i.e., in a hospital or patient home).
With these precautions, which a responsible user can easily follow, protection can be ensured against the malicious replacement of a command using a compromised reader, or against an attacker sending a malicious command him/herself in order to frame said user.
Due to the above risk-based, multi-factor authentication, a user cannot falsely deny his/her involvement in a certain implant access because the alternative explanation implies that (1) the attacker stole a valid reader, card and pin, (2) accessed the implant from within the hospital and during the user's working hours, and (3) $R$ and $\doctor$ were not reported as stolen. The combined probabilities of all these events occurring at once is extremely small, or, in other words, the non-repudiation gap is effectively bridged by the introduction of above checks.

\subsubsection{Bedside-reader operation}

The online mode also facilitates bedside-reader operation (see Fig.~\ref{fig:imd-emv-system}). Here, only the CPP and DH-based handshake between the bedside $R$ and $\server$ (from reader-card authentication phase), the session-key establishment phase, and the main phase (with a few differences, as indicated in Fig.~\ref{fig:kerberos} and Fig.~\ref{fig:main-phase}, respectively) need to be executed, since the commands and responses are only sent and read by $\server$.
Moreover, since the remote monitoring done in practice is only \textit{read only}, i.e., with the lowest access privileges, there is no need for non-repudiation \textit{if} the read-only access control is implemented correctly.
This can be done if $sig$ in step 6 is replaced by MAC of CMD from $\server$ (i.e., $\mathsf{MAC}_{K_{SI}}(CMD, N_R, N_I)$).
Using this MAC, $I$ is able to verify that the command came from the server, and hence, it can be executed with read-only privileges.
Finally, the hospital staff can retrieve the critical treatment data by logging into $\server$.
It can be argued that this remote-access mode should support \textit{read/write} access instead of just read-only in order to enable remote firmware updates.
However, we stress that such updates should always occur in the presence of a qualified professional.
This is important in case patient health suddenly deteriorates due to the update process.
Moreover, in practice it is quite common and acceptable to get the IMD firmwares updated at the clinic in the presence of a physician~\cite{fda2019abbott}.
This mode is also useful for securely retrieving the stored signatures pertaining to previous programming sessions in order to free up limited IMD memory.

\subsubsection{IMD access from a non-local location}


\Figure[!t]()[trim={0.5cm 3.3cm 0.5cm 3.3cm},clip,scale=0.34]{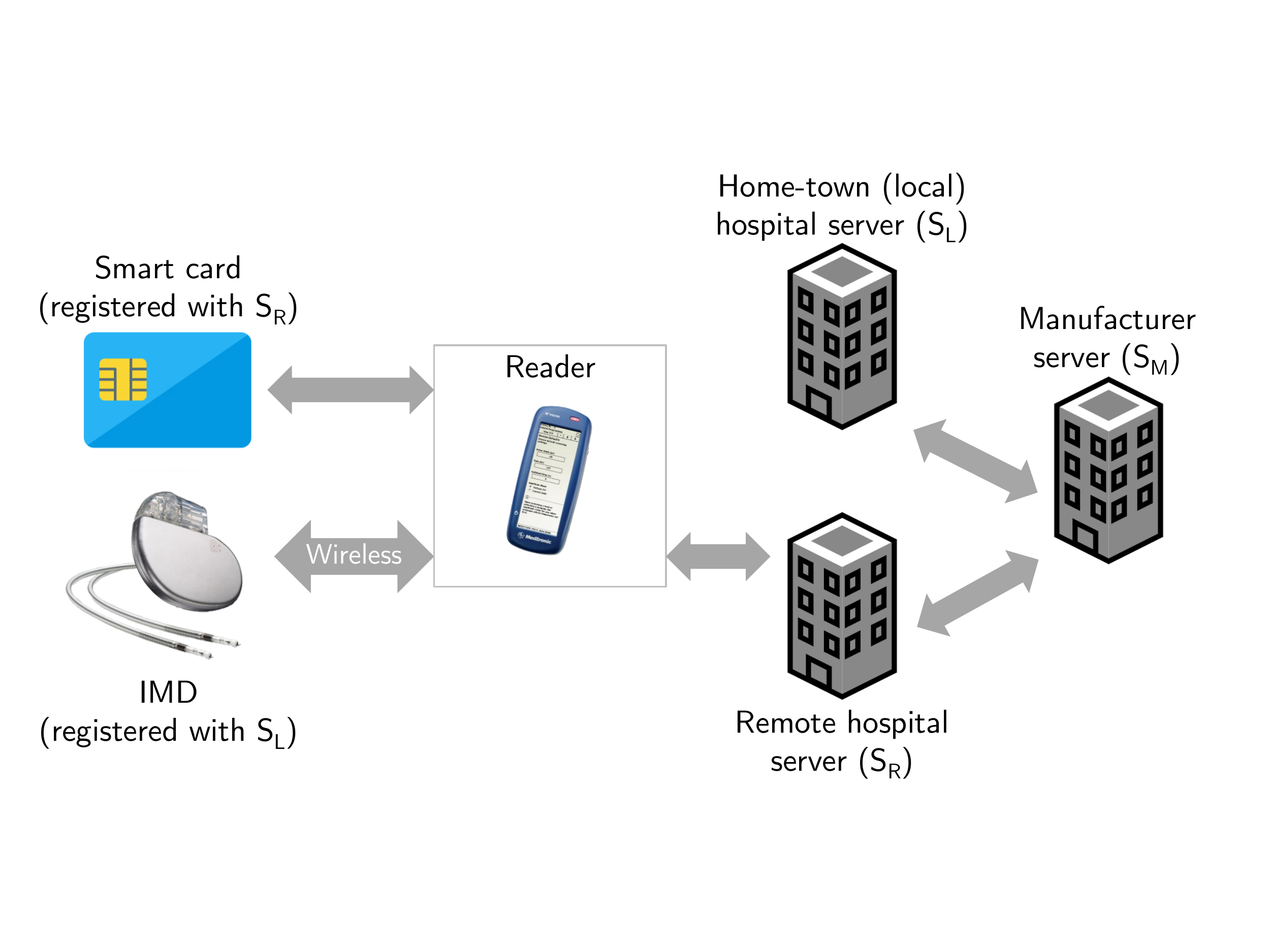}
{Scenario when the patient is out of town\label{fig:imd-emv-system-remote}}

In Section~\ref{sec:protocol_interfaces}, we discussed that $\doctor$ and $I$ are registered at the local hospital ($\server_L$), or in other words, they share their respective symmetric keys with the hospital server.
During travels or when the patient is out of town, a situation may arise that requires access to the IMD for status monitoring. In this case, the scheme from Fig.~\ref{fig:imd-emv-system}, can still work if the patient is in possession of $\reader$ and his/her $\doctor$.
However, for treatment updates, which require higher access privileges, the patient would need to visit a nearby (remote) hospital ($\server_R$). In this case, the above scheme would not work straightaway since the IMD is not registered at $\server_R$ and the remote-location physician's $\doctor$ is not registered at $\server_L$.
Hence, minor extensions are required (see Fig.~\ref{fig:imd-emv-system-remote}), in which $\server_R$ establishes a secure connection with $\server_L$ via an IMD-manufacturer server $\server_M$.
$\server_M$ maintains a list of all the IMDs in service and the hospitals at which they are registered.
Based on $ID_I$ sent by $R$ to $\server_R$ (and then $\server_R$ to $\server_M$) during the session-key establishment phase (see Fig.~\ref{fig:kerberos}), $\server_M$ determines $\server_L$ and establishes a secure connection with it.
$\server_R$ sends $K'_{RI}$, the relevant identifiers, nonces and $P_{\doctor}$ to $\server_L$ (via $\server_M$) so that $\server_L$ is able to construct $m_I$ and send it back to $\server_R$.
The protocol then proceeds normally and the IMD eventually retrieves $K'_{RI}$ after decrypting $m_I$.

\subsection{Offline mode}
\label{sec:offline-mode}


\Figure[!t]()[trim={0 0.5cm 0 1.3cm},clip,scale=0.61]{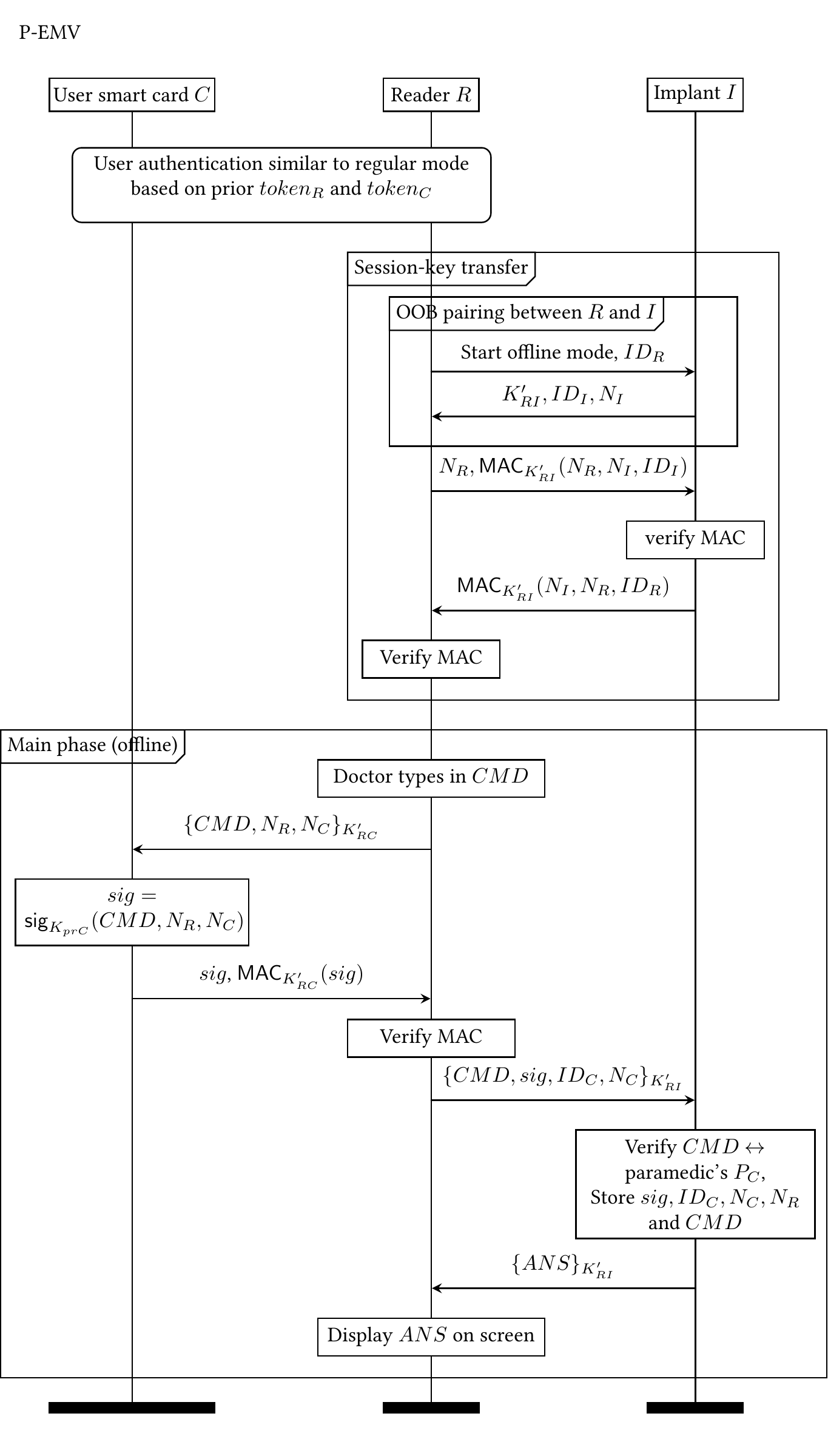}
{\simdprotocol (Offline mode)\label{fig:emv-protocol-offline}}

In the absence of an active Internet connection and hence, the TTP ($\server$), e.g., during emergencies, $R$ and $I$ need to establish a \textit{temporary} shared key so that they can communicate directly in a secure manner.
We propose to employ an OOB-channel-based key exchange while using the principle of \textit{touch-to-access} (as discussed in Section~\ref{sec:related-work}).
This principle is employed by $I$ to establish trust with $R$ since we assume $R$ to be untrustworthy from the perspective of the IMD.
We propose to either use \textit{ultrasound communication} or \textit{galvanic coupling} as the OOB channel (between $R$ and $I$) since they result in virtually zero information leakage compared to other coupling methods, such as capacitive coupling~\cite{tomlinson2018comprehensive}.
Moreover, they have an advantage over biometric-based touch-to-access mechanisms (mentioned in Section~\ref{sec:related-work}) in that they do not require any initial RF communication messages before the IMD is sure that the external entity is in close proximity. This provides an additional security layer, which is critical for the pre-deployment configuration that will be discussed in Section~\ref{sec:without-nr}.

Assuming that galvanic coupling is used, the paramedic places the OOB interface of the reader on the patient skin\footnote{Touching the skin is mandatory for the galvanic channel to function.} at a point that is nearest to the IMD.
The patient is assumed to thwart advances of a stranger trying to place a reader on his/her skin, if there is no emergency or a need for treatment. Hence, the implant assumes that the message received from the OOB interface is from a trustworthy source.
In other words, in offline mode, the IMD-system security hinges on this OOB pairing and favors availability over security but in a more controlled fashion than state of the art.

The protocol is shown in Fig.~\ref{fig:emv-protocol-offline}.
The paramedic is required to perform reader-card authentication when starting his/her duty, so that both $R$ and $\doctor$ obtain their respective tokens from $\server$.
When IMD access is required in an offline setting, $R$ first initiates user authentication with the paramedic smart card in the same way as in the regular mode.
During user authentication, $R$ verifies that its internal real-time-clock value is less than $\sTL$.
Through the OOB channel, $R$ sends a request for offline access along with its identifier.
Upon receiving this request, the implant assumes that this is an offline scenario since this channel is activated only in such extraordinary circumstances.
As a result, it generates a random key $K'_{RI}$ and its nonce and sends them along with $ID_I$ to the reader using the same channel.

$R$, then, initiates session-key confirmation with $I$ in which both entities verify each other's MACs that are generated using $K'_{RI}$.
In order to update or inquire about the implant operation, the paramedic enters the command on the reader interface, which is encrypted using $K'_{R\doctor}$ and is sent to $\doctor$. The card digitally signs this command and sends it back to $R$.
$R$ encrypts the command using $K'_{RI}$, calculates its MAC and sends it to $I$ along with $\mathsf{sig}_{Kpr\doctor}(CMD, N_R, N_{\doctor})$.
This signature and $CMD$ are stored by the IMD and are required to ensure non-repudiation, as already discussed in Section~\ref{sec:regular_mode}.
The IMD responds with an answer encrypted by the same session key, which is subsequently displayed on the reader display.
The session key expires in a manner similar to that in the regular mode.

In offline mode, the user is only allowed paramedic-level privileges, which have less access rights compared to a technician (see Section~\ref{sec:requirements}).
The use of the OOB channel makes it straightforward for the IMD to decide on granting only paramedic-role commands.

\subsubsection{Offline access with/without non-repudiation and access control}
\label{sec:without-nr}

We also propose a second flavor of the offline mode in which non-repudiation and user authentication are not a requirement. This is suitable for less critical implants, such as neurostimulators.
This flavor does not require a smart card, and as a result we do not require the reader-card- and user-authentication phases in addition to signature generation.
This improves \textit{usability}, since the paramedic is not required to perform reader-card authentication when starting their duty.
In this scheme, the touch-to-access principle is deemed to be sufficient in order to ensure trust establishment.
It is important to note that, for \simdprotocol, supporting non-repudiation during offline mode has to be decided before IMD-system deployment since it cannot be configured at runtime, so as to avoid exploitation.

\subsubsection{Offline access with/without reader-interface standardization}
\label{sec:without-standardization}

As indicated in Section~\ref{sec:requirements}, supporting emergency access in the field requires a standardized reader interface, which demands collaboration between major IMD manufacturers.
In order to facilitate this \textit{\textbf{multi-manufacturer environment (\sMulti)}}, there has to be one agreed-upon root CA that grants certificates to the manufacturers, who can then act as intermediate CAs that sign public keys of $\server$, $R$ and $\doctor$.
As things stand, however, \textit{true} emergency access does not exist in commercial IMDs.
As long as this remains an open issue, the above standardization is not required, and as a result, \simdprotocol can be simplified by eliminating the need for a global root CA.
Emergency-access support in \simdprotocol is intended to be there in anticipation of any future changes in this regard.

\subsection{Summary of protocol configurations}

The different configurations of \simdprotocol are highlighted in Fig.~\ref{fig:container-diagram}. The dotted boxes indicate (fixed) pre-deployment configurations, which cannot be changed at run-time. Such configurations were discussed in~\Cref{sec:without-nr,sec:without-standardization}.

\simdprotocol is designed in such a way that an attacker cannot target one mode over another for exploitation. For instance, the offline mode is only triggered after an OOB access, which is protected by the touch-to-access principle. Moreover, the sub-modes of online access only come about by disabling certain \simdprotocol steps instead of switching to a totally independent behavior.


\Figure[!t]()[trim={3.5cm 5.2cm 3.5cm 5.2cm},clip,scale=0.45]{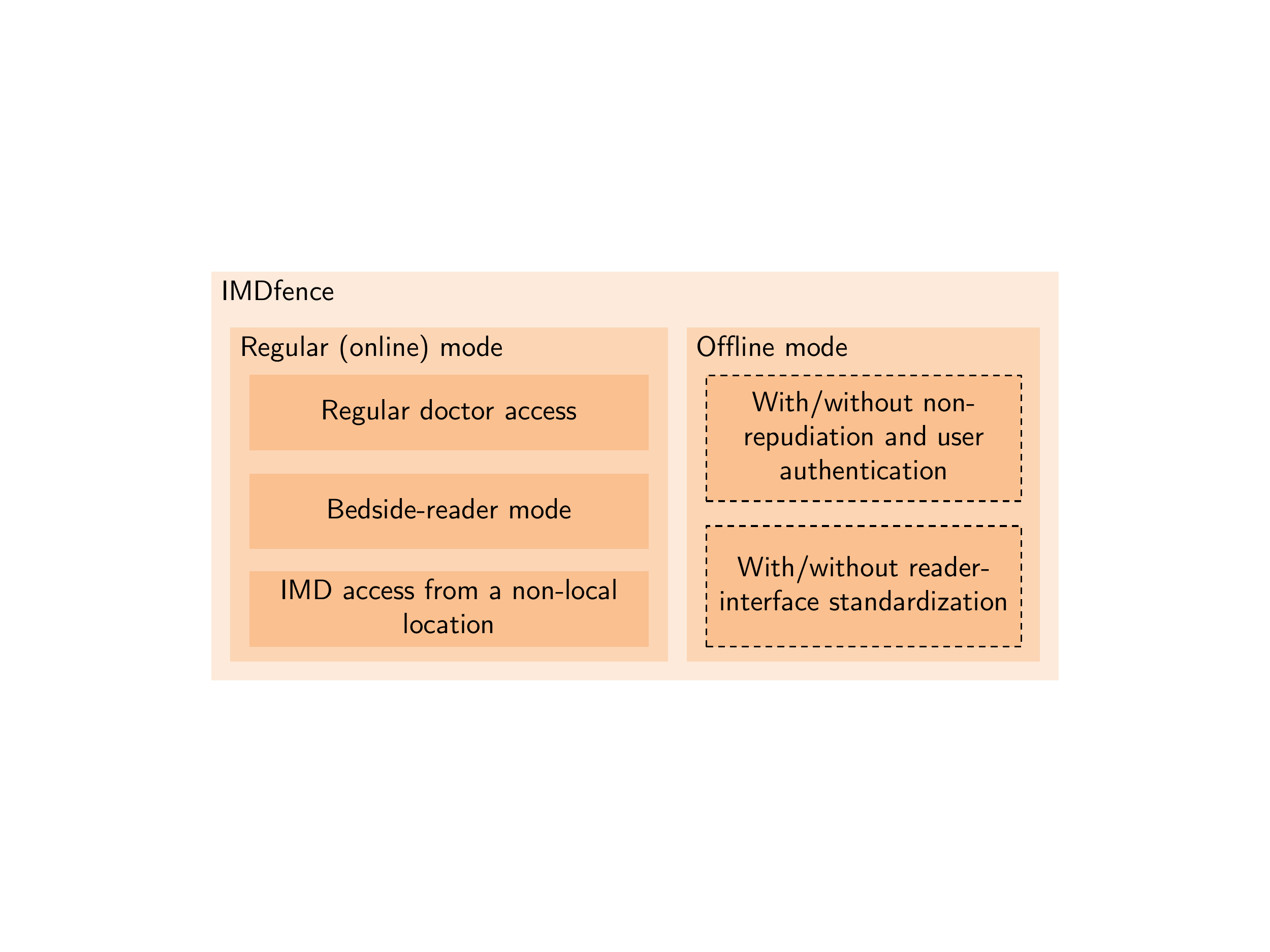}
{\simdprotocol configurations and use cases\label{fig:container-diagram}} 

\section{Evaluation}
\label{sec:evaluation}

In this section, we evaluate our system in terms of security feasibility and also look into the handling of battery-DoS protection for IMDs.

\subsection{Security analysis}
\label{sec:security-analysis}

\subsubsection{Automatic validation using AVISPA tool}

For the automated and formal validation of \simdprotocol, we used AVISPA (Automated Validation of Internet Security Protocols and Applications)~\cite{avispa2006}.
Any protocol to be validated using this tool is specified using the High-Level Protocol Specification Language (HLPSL).
An HLPSL specification consists of a description of the \textit{principals} (i.e., $R$, $I$, $\doctor$, $\server$ and the user in our case), security \textit{goals} of the protocol, and the details of the \textit{session(s)} to be analyzed.
AVISPA integrates four back-end engines that provide different types of automatic analysis of an HLPSL specification~\cite{avispa2006}.
The tool helps in detecting vulnerabilities against Man-in-the-middle and replay attacks. It also detects whether the HLPSL specification is executable, i.e., all the specified protocol states are traversable.
Using AVISPA, we can also optimize our protocols by removing certain parameters from the messages in order to reduce communication overhead and analyze if this results in a new vulnerability.

The analysis of \simdprotocol using AVISPA is summarized in Table~\ref{table:avispa}. The handshake-specific protocol requirements (\sConf, \sNR, \sAC and \sAuth) are satisfied by specifying the appropriate goals.
In phase III, $\server$ extracts user privileges from $Cert_{\doctor}$ after successful authentication of $\doctor$, based on $N_{\server}$ in $m_{\server\doctor_2}$. $I$ then verifies $\server$ based on $N_I$ to complete the chain from the card to the implant in order to ensure access control.
In order to check non-repudiation using the tool, the server verifies that the retrieved $sig$ from the IMD originated from $\doctor$ during the session corresponding to $N_{\server}$.

\begin{table}
	\centering
	\begin{threeparttable}
	\caption{Summary of AVISPA analysis}
	\label{table:avispa}
	\small
	\begin{tabular}{lll}
		\toprule
		\
		Phase & AVISPA goal\tnote{*} & Coverage \\
		\midrule
		I. Reader-card auth.  		& Secrecy of $K'_{R{\doctor}}$ & \sConf, \sAuth \\
									& $\doctor \rightarrow R \; | \; N_{\doctor}$ &   	\\
									& $R \rightarrow \doctor \; | \; N_{R}$ &   	\\
									& $\server \rightarrow \doctor \; | \; N_{\server}$ &   	\\
									\hline
		II. User auth. 				& Secrecy of $PIN$ & \sConf, \sAuth  \\
									& $\doctor \rightarrow U \; | \; PIN$ &   	\\
									\hline
		III. Session-key est.		& Secrecy of $K'_{RI}$ & \sConf, \sAC,  \\
									& $\server \rightarrow \doctor \; | \; N_{\server}$ & \sAuth \\
									& $I \rightarrow R \; | \; N_I$ &   	\\
									& $R \rightarrow I \; | \; N_R$ &   	\\
									& $I \rightarrow \server \; | \; N_I$ &   	\\
									\hline
		IV. Main Phase 				& Secrecy of $CMD, ANS$ & \sConf, \sNR  \\
									& $\server \rightarrow \doctor \; | \; N_{\server}$ &  \\
		\bottomrule
	\end{tabular}
	\begin{tablenotes}
	\footnotesize
	\item [*] $A \rightarrow B \; | \; N$: $A$ authenticates $B$ based on value $N$ \\
	Secrecy of $N$: Confidentiality of value $N$ is ensured
	\end{tablenotes}
	\end{threeparttable}
\end{table}

\subsubsection{Reader-specific attacks}
\label{sec:attack-scenarios}

When considering all possible attack scenarios, we define the following reader types:
\begin{enumerate}
	\item \textit{Valid $R$ ($\valid$):} This is a legitimate device, which is \textit{not} reported as stolen.
	\item \textit{Stolen $R$ ($\stolen$):} A legitimate device which is \textit{reported} as stolen.
	\item \textit{Hacked $R$ ($\hacked$):} A stolen reader which is also modified by $A$ in order to e.g., replace the signature or $CMD$.
	\item \textit{Forged $R$ ($\forged$):} A custom-built or software-defined radio used by $A$ in order to communicate with an implant. This reader does not have any pre-shared keys with $\server$.
\end{enumerate}

The following scenarios are possible in terms of user-reader combinations (which are also summarized in Table~\ref{table:scenarios}):

\begin{table}
	\centering
	\begin{threeparttable}
		\caption{Enumeration of attack scenarios Sn in terms of user-reader combinations}
		\label{table:scenarios}
		\small
		\begin{tabular}{lllll}
			\hline
			& \multicolumn{4}{c}{Reader} \\\cline{2-5}
			&   Valid   & Stolen & Hacked & Forged  \\
			\hline
			Trusted, honest user  		&   \leguserValid   & \anyuserStolen & \leguserForged & \leguserForged \\
			Trusted, malicious user		&   \leguserValid   & \anyuserStolen & \maluserHacked & \maluserForged  \\
			Attacker 			  		&   \attackerValid   & \anyuserStolen & \attackerForged & \attackerForged  \\
			\hline
		\end{tabular}
	\end{threeparttable}
\end{table}

\textbf{\leguserValid~-- Any user \& \boldmath$\valid$:} This is the most common scenario, which must be handled by \simdprotocol.
$A$ cannot insert a false signature remotely (in order to frame someone) since the connection between $R$ and $\doctor$ is protected by MAC-based integrity checks.
Moreover, an insider attack (from a legitimate, malicious user) should be detected by the non-repudiation check.
However, after sending a malicious command, such a user can attempt multiple harmless write commands in order to eventually overwrite the signature corresponding to the malicious command. We term this as the \textit{signature-overwrite attack}.
For each command, 72 bytes of flash space is required to store the signature and the associated session parameters. As an example, if a 32-kB flash memory is allocated for signature storage, 456 attempts will be required to successfully overwrite the targeted signature, which is highly impractical. Even if the user manages to achieve this, the signature record will still point to an abnormally high number of write commands corresponding to a single session, which will raise suspicions.

\textbf{\anyuserStolen~-- Any user or attacker \& \boldmath$\stolen$:} No individual will be able to use $\stolen$ because of the checks involved in the reader-card-authentication phase.

\textbf{\leguserForged~-- Trusted, honest user \& \boldmath$\hacked$/$\forged$:} In order to frame someone, $A$ has to force the legitimate user to use a hacked reader, which replaces the command with an incorrect one.
As a guideline, $R$ must be issued from a trusted repository, which rules out the use of $\hacked$ and $\forged$ for trusted users.

\textbf{\maluserHacked~-- Trusted, malicious user \& \boldmath$\hacked$:} Legitimate malicious users can cover their tracks by using a hacked reader that can replace the signature corresponding to a malicious command, which is to be stored in the IMD, with the one corresponding to a safe command. Such an attack is quite costly to execute and is time-critical since it will involve colluding with someone who has advanced engineering skills while requiring that $\hacked$ is not reported as stolen.
Since, the user is considered trusted by the patient and can thus be in close proximity, he/she has far easier and inexpensive means to harm the patient without getting caught.

\textbf{\maluserForged~-- Trusted, malicious user \& \boldmath$\forged$:} Such a user cannot send commands using a forged reader in an online case since $\forged$ does not share a key with $\server$. In the offline case, however, such a user can use a forged reader that is able to create a bogus $sig$ and hence does not require any involvement of $\doctor$. Moreover, he/she can use the OOB-pairing interface because of being considered as trusted by the patient. Similar to \maluserHacked, such a scenario also requires hiring an advanced attacker to develop such a reader, and based on the touch-to-access assumption, the user has significantly easier methods to harm the patient.
	
\textbf{\attackerValid~-- Attacker \& \boldmath$\valid$:} For online access, the security protocol will break if $A$ gets hold of a valid reader, card and its associated PIN, accesses the IMD from within the hospital and during the user's working hours, \textit{and} $\doctor$ is not reported as stolen.
It is recommended that the user protects her card and PIN, or immediately reports it in case it is lost.
Moreover, as a guideline, the user should never lend or sell $R$ to a third party.
The protocol will also break if $A$ gets hold of an \textit{OOB-paired} reader and a card with valid respective \textit{tokens}, \textit{and} knows the PIN. We assume that the paramedic resets the pairing after treatment.
Overall, $A$ cannot effectively launch the above attacks since the likelihood of all the dependencies being true is extremely low.

\textbf{\attackerForged~-- Attacker \& \boldmath$\hacked$/$\forged$:} For online access, $A$ will not be able to use $\hacked$ because of the reasons mentioned in \attackerValid above. Similarly, $A$ cannot use $\forged$ since it does not have a shared key with $\server$.
Moreover, for an offline scenario, getting hold of these readers will not help an attacker $A$ since the main symmetric key ($K_{RI}$) comes from $I$ in the OOB pairing process. Hence, to gain advantage using these readers, $A$ would still need to get close to $I$ (touch-to-access).

\subsubsection{Smart-card-specific attacks}
\label{sec:emv_attacks}

Since \simdprotocol employs smart cards, it is important to ensure that it is safe from the weaknesses~\cite{van2014security, van2016emv} present in another widely used smart-card system: EMV (Europay, Mastercard, and Visa).
These vulnerabilities exist due to the availability of less secure options for backward compatibility and due to a problematic threat model, in which the reader (i.e., the POS terminal) is assumed to be uncorrupted.

One major issue is that most of the important data is exchanged in plain-text (e.g., account data, amount etc.) since the terminal and the card do not share a symmetric key.
Moreover, in the offline use of the cards that do not support public-key cryptography, the PIN is also sent as plain-text.
An attacker can modify the unencrypted initialization messages to force the terminal to use this mode~\cite{van2016emv}.
The PIN can be recorded using e.g., a hacked terminal that has additional probes to read data from the smart card interface.
In case of an offline-encrypted PIN, the terminal can be hacked to record the keystrokes.
Using the account data and PIN, the attacker can create a magnetic-strip card for use in a country that does not support chip-based smart cards~\cite{adida2006phish}.

Another issue is that the terminal cannot use MAC to authenticate messages from the card since they do not share a symmetric key. Cards following the Combined-Data-Authentication (CDA) scheme from EMV address this by employing signatures.
However, in the schemes prior to CDA, the terminal is unable to verify the authenticity of all the card messages either due to unavailability of signatures (in the case of Static Data Authentication, SDA) or the signature-less transaction messages (in the case of Dynamic Data Authentication, DDA).
As a result, an SDA card can be cloned for use in offline transactions~\cite{van2016emv}, and a stolen DDA card can be employed in a \textit{two-card attack}, in which the attacker uses his/her own card for PIN verification and uses the stolen card in the transaction phase~\cite{anderson2008security}.
Moreover, the card response at the end of PIN verification is unauthenticated. As a result, this response can be modified to deceive the terminal into assuming that the entered PIN is correct.

All these attacks exist because in EMV some of the critical data is left unencrypted or not signed.
In contrast, in both the online and offline modes of \simdprotocol, all data between $R$ and $\doctor$ is encrypted and is authenticated using MACs.
Additionally, our recommendation to avoid magnetic-strip-based cards rules out cloning. Similarly, avoiding contactless cards removes an additional attack vector.

Another far more advanced type of attack is the \textit{relay attack}~\cite{adida2006phish,van2014security}, which exploits the fact that the card users cannot know for sure if the display of the terminal is showing correct information.
It is a time-critical attack where two transactions are simultaneously taking place.
The victim inserts his/her card in a counterfeit terminal (e.g., at a restaurant), which is connected to a fake card of the attacker that is inserted in a valid terminal (e.g. at a jewelry store).
The details of the fraudulent transaction are forwarded to the victim's terminal. Her screen shows the correct information, but in effect she pays the amount for the other party.

We observe that the relay attack is far less likely in the case of \simdprotocol since it requires a legitimate user operating a forged reader. This corresponds to scenario \leguserForged discussed in Section~\ref{sec:attack-scenarios}.

\subsubsection{Selection of $\sTL$}
\label{sec:t_and_nac}

The touch-to-access principle guarantees that an unreasonably high $\sTL$ (reader-card-authentication lifetime) value does not cause a security vulnerability in \simdprotocol, as evident from Section~\ref{sec:attack-scenarios}.
However, the careful reader may have noticed that a prolonged offline operation enabled by such a large value may result in $R$'s and/or IMD's firmwares becoming outdated.
On the other hand, a very small value hinders legitimate access, i.e., availability.
Therefore, the hospital server should ensure that $\sTL$ is assigned an appropriate value (within maximum and minimum limits) based on the patient's location and the reader-IMD usage patterns.

Regarding the patient's locality, the probability of having stable Internet connectivity is higher when the patient is based in an urban area compared to a rural setting.
Moreover, it stands to reason that the chances of attacker presence ought to be higher in an urban environment.
Hence, it makes sense to assign a lower $\sTL$ value for urban areas compared to rural environments.
When assigning the $\sTL$ value, reader-IMD usage patterns should also be taken into consideration, which depend on the patient condition and IMD type, ranging from critical implants, such as cardiac defibrillators, to less critical ones, such as neurostimulators. The IMDs requiring frequent reader access should be granted a larger $\sTL$ value.
Further investigation on this topic is interesting but is considered outside the scope of this work.

It should be noted that the (re)setting of $\sTL$ can be performed throughout the operational lifetime of the IMD. The physician is required to manually modify this parameter (in $\server$) based on the above guidelines, which then ultimately take effect in the reader-card authentication phase (see Fig.~\ref{fig:initial_kerberos}).

\subsection{Availability -- DoS protection}
\label{sec:availability}

As highlighted in Section~\ref{sec:requirements}, one of the system requirements is to ensure that the IMD is always \textit{available} for treatment. One high-likelihood and low-cost attack that affects this requirement is the battery-DoS attack, as practically demonstrated in~\cite{halperin2008pacemakers,marin2016security}.
This attack forces the IMD to continuously run energy-consuming operations, which results in battery depletion and ultimately causes device shutdown.
For example, the attacker can repeatedly try to establish a connection with the implant using incorrect credentials.
The IMD will scrutinize each invalid request through energy-consuming authentication operations, which will drain its battery despite failing to authenticate properly.

The IMD can defend against battery DoS by employing a \textit{zero-power defense} (ZPD) scheme in which the authentication operation is executed using borrowed energy~\cite{halperin2008pacemakers}.
This energy can be harvested from the incoming RF communication messages from the external reader.
The IMD switches to battery power only after it has successfully authenticated the external entity.

Another type of DoS attack can occur when the attacker sends repeated communication requests to the implant.
For an IMD with a single-processor, such requests may block the device from performing its primary medical functionality.
To protect against this, a dual-CPU paradigm can be employed, in which the first CPU executes the original medical functionality, while the second CPU is responsible for dealing with the (secure) communication requests. This dual-core organization offers, then, both functional and power decoupling, which effectively shields the IMD main functionality from battery-DoS attacks, as previously showcased in~\cite{strydis2013system}.

In order to assess the viability of \simdprotocol under energy-harvesting conditions (be it in single- or dual-CPU configuration), we construct the following experimental setup:

\noindent\textit{(I) Computational costs:} Similarly to~\cite{siddiqi2019towards}, we employ an ARM Cortex-M0+ based 32-bit MCU~\cite{tinygecko}. Due to its ultra-low-power capabilities, and the on-board hardware-accelerated, security building blocks (i.e., encryption, MAC, hash function, random-number generator etc.), this MCU is becoming increasingly employed in IoT and WBAN settings~\cite{businesswire2015}, and hence, is a plausible choice for this evaluation.
The security-related computations, i.e., authenticated encryption (AES-128), cipher-based MAC and random-number generation were performed using the MCU's dedicated peripherals (``CRYPTO'' and ``TRNG''); thus, in our energy measurements, hardware-accelerated primitives are considered. However, as a reference, we also include a software-only MCU implementation of \simdprotocol.

\noindent\textit{(II) Wireless-communication costs:} Commercial transceiver ZL70103 specifically designed for IMDs has been used~\cite{microsemi70103}. To get reasonable energy costs for (encrypted) data transmission, we chose packet-size lengths similar to the ones used in low-cost RFID tags, due to their similarities with IMDs in terms of computational, memory and energy constraints~\cite{strydis2013system}.
Hence $N$, $ID$, $CMD$ and $ANS$ were set to 32, 96, 32 and 64 bits, respectively.
The $sig$ size was set at 384 bits, which corresponds to an ECDSA (Elliptic-Curve Digital-Signature Algorithm) signature with a 96-bit security level.

The protocol sequence executed by the IMD is shown as numbered steps in~\Cref{fig:kerberos,fig:main-phase}.
In the case of hardware-accelerated primitives, the energy consumption for these steps is shown in Fig.~\ref{fig:plot_emv_imd} using a supply voltage of 3.3~V, and the default MCU and transceiver clock frequencies of 19 MHz and 24 MHz, respectively.
The transceiver data rate is set at 400 kbps (with an effective rate of 265 kbps).
We observe that the energy required for authentication ($E_{auth}$), i.e., for steps 1 to 4 in Fig.~\ref{fig:kerberos}, is only 59.6 $\mu$J.
In the case of software implementation, however, $E_{auth}$ is only 119.4 $\mu$J, as shown in Fig.~\ref{fig:plot_emv_imd_sw}.
For such a low harvested-energy requirement ($E_{auth}$), it has been demonstrated before in~\cite{siddiqi2019towards} that real-time performance is possible in the IMD with or without hardware acceleration.
Total IMD energy consumption per type of activity is also shown Fig.~\ref{fig:plot_emv_imd_total}.


\Figure[!t]()[trim={0.4cm 0.45cm 0.45cm 0.4cm},clip,scale=0.35]{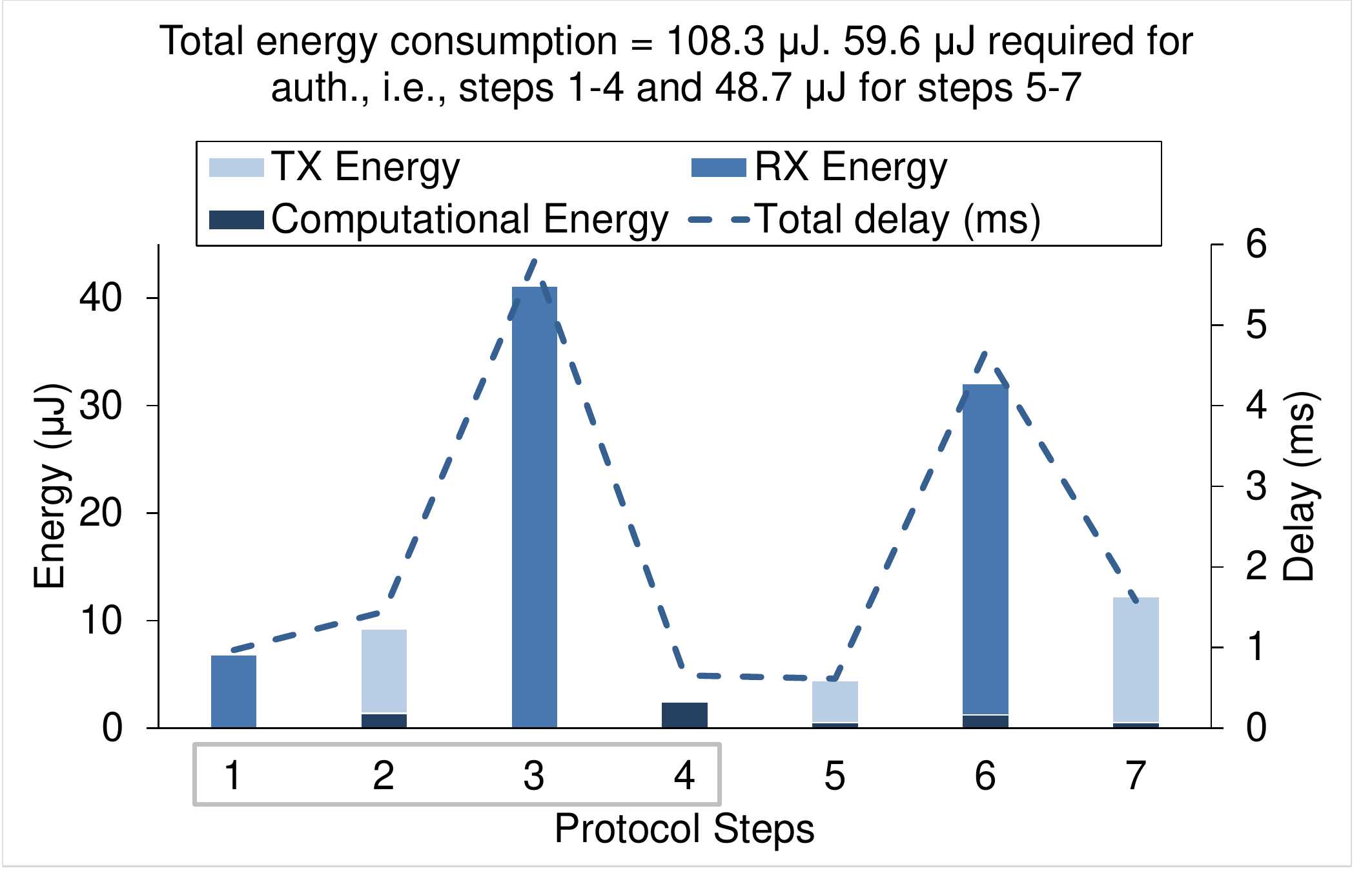}
{IMD energy consumption and performance per \simdprotocol -protocol step while using hardware-accelerated security primitives \label{fig:plot_emv_imd}}


\Figure[!t]()[trim={0.4cm 0.45cm 0.45cm 0.4cm},clip,scale=0.35]{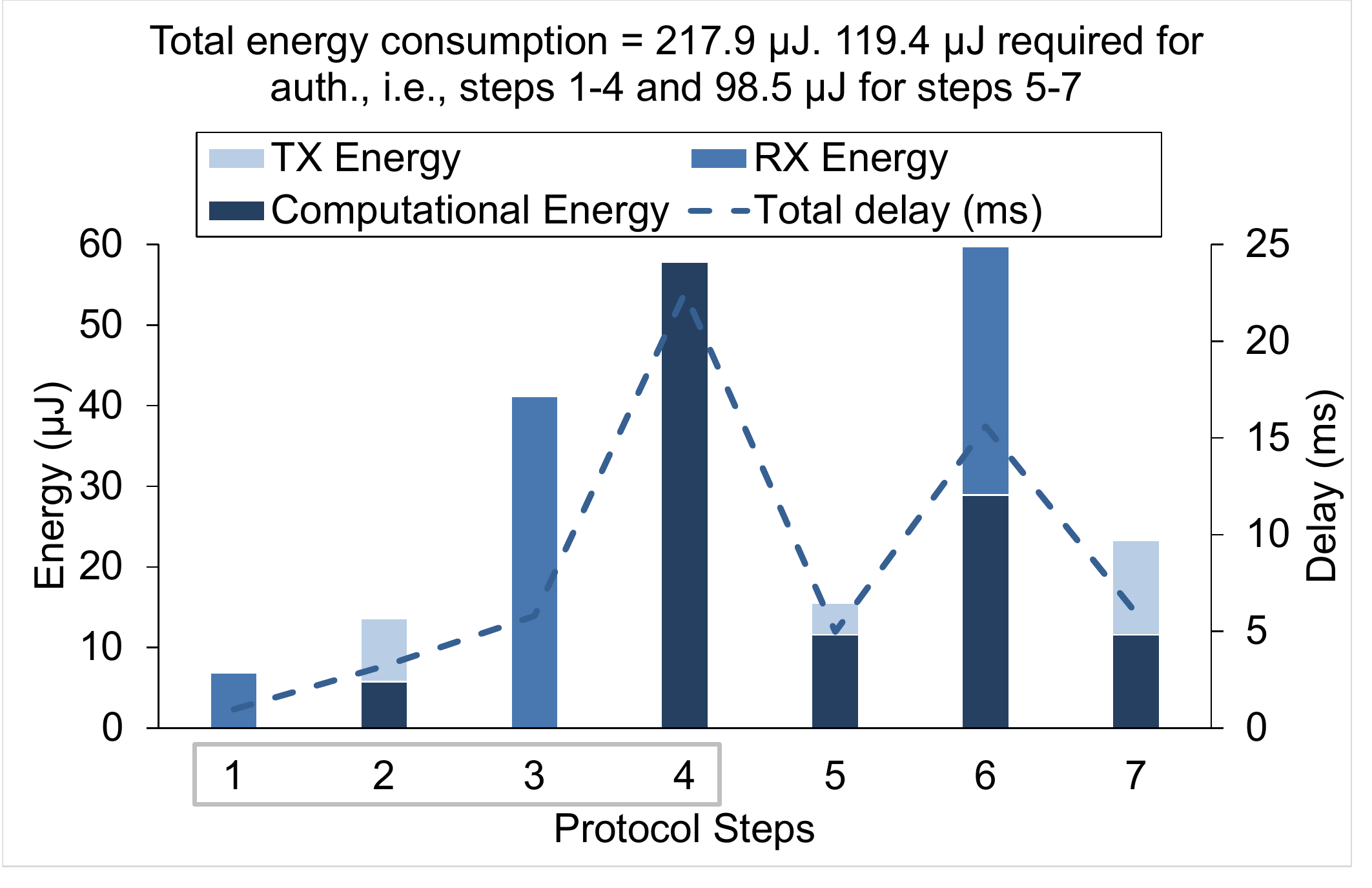}
{IMD energy consumption and performance per \simdprotocol -protocol step when implementing the security primitives in software \label{fig:plot_emv_imd_sw}}


\Figure[!t]()[trim={1.0cm 4.9cm 1.0cm 0.3cm},clip,scale=0.35]{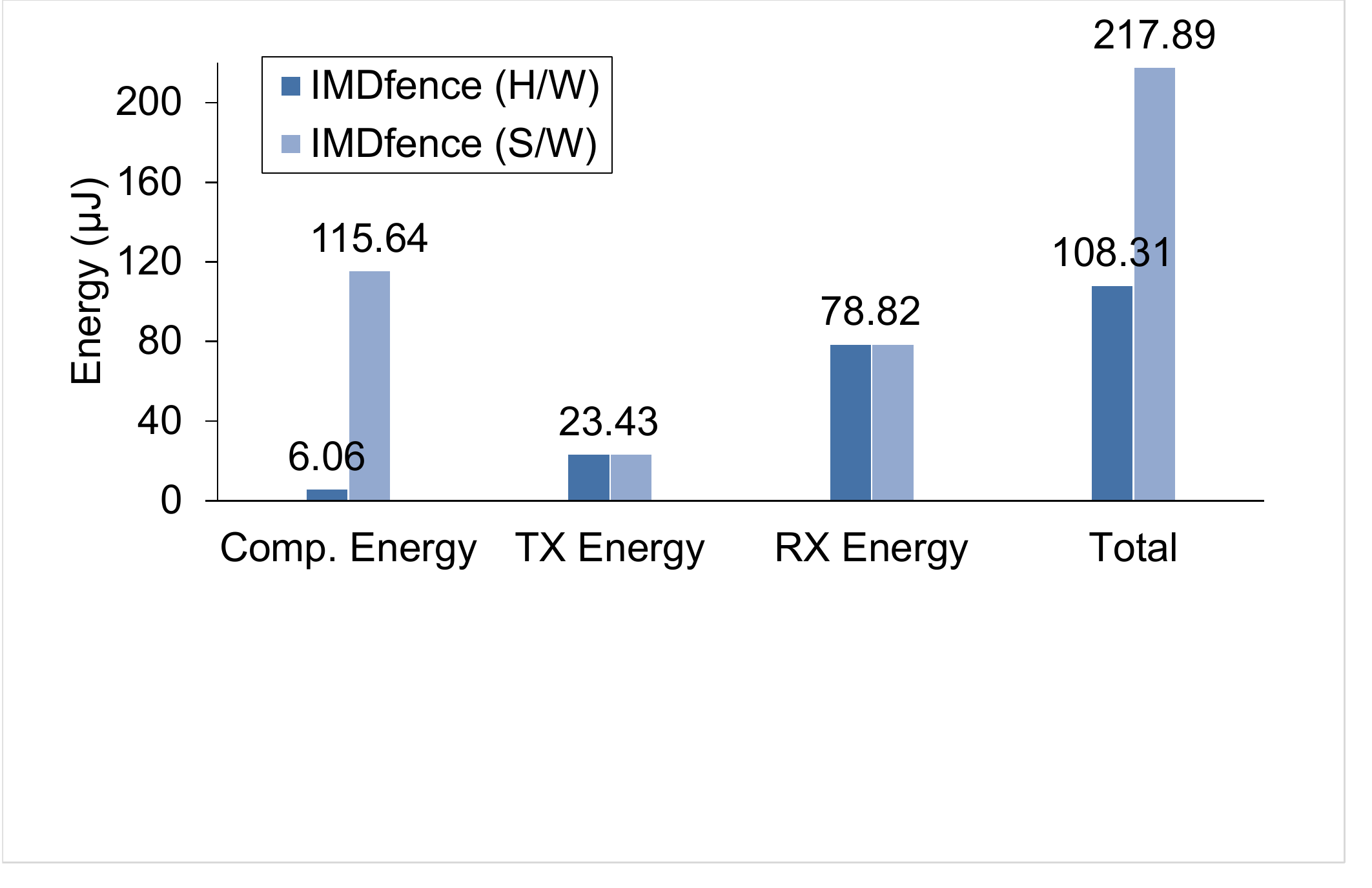}
{IMD energy consumption per \simdprotocol -protocol activity\label{fig:plot_emv_imd_total}}

\subsection{IMD lifetime}
\label{sec:imd-lifetime}

In the previous section, we discussed the feasibility of \simdprotocol under energy-harvesting conditions to defend against battery-DoS attacks. In this section, we wish to assess the total energy costs that the \simdprotocol protocol incurs over the whole lifetime of a modern IMD. To do so, we need to consider realistic usage patterns of actual devices, drawn from medical practice.
There are two prominent IMD classes: neurostimulators and cardiac implants. Neurostimulators typically consume more power than cardiac devices~\cite{mehta2018when} and, therefore, often come with rechargeable batteries which would pose no challenge for \simdprotocol. Cardiac implants, on the other hand, are not rechargeable due to their critical nature~\cite{siddiqi2019towards}, and represent more \textit{pessimistic} devices to assess \simdprotocol against. Thus, for our evaluation here, we consider a communication session between a pacemaker and a commercial bedside reader (Merlin@home\textsuperscript{TM})~\cite{merlin2015faq}.

We consider different data volumes being transferred between the reader and IMD, ranging from a daily two-minute\footnote{This corresponds to an unencrypted session. An equivalent secure session (by employing \simdprotocol) will take longer than two minutes due to the additional data transferred.} communication session to a two-minute weekly session.
Since this reader is intended for monitoring the IMD status, it is assumed that most of the communicated data is transferred from the implant to the reader (e.g., in the form of data logs).
Hence, the size of $ANS$ is increased from 64 bits (for a basic session) to roughly 3 MB in order to form a two-minute session.
However, for worst-case analysis, the transceiver is considered to be enabled throughout this session and we do not assume the use of energy harvesting for ZPD.
Moreover, without loss of generality and in order to more accurately (and pessimistically) quantify the cost of adding \simdprotocol to an existing system, we consider a dual-CPU IMD, as discussed in the previous section. In this configuration, the security CPU is assumed to execute the complete \simdprotocol protocol, while the medical CPU is set to a 5\% duty cycle (active vs. sleep mode), based on typical pacemaker usage~\cite{lindqvist2005compression}, and consumes 20 $\mu$J per heartbeat to provide electrical-stimulation impulses, based on reported figures of commercial devices~\cite{deterre2013toward}.

With the above consideration, the impact of \simdprotocol on IMD-battery lifetime can be visualized using Fig.~\ref{fig:lifetime-boxplot} for different implantable-grade battery sizes~\cite{eaglepicher}.
The variability in each data point captures the different volumes of data transfer between the reader and IMD.

Since the majority of the cryptographic operations in the protocol (authenticated encryption and MAC) are based on symmetric block ciphers, as shown in~\Cref{fig:kerberos,fig:main-phase}, it is very interesting to investigate the impact of different cipher versions and/or implementations thereof on IMD lifetime, e.g., a pacemaker. More box plots have, thus, been added to Fig.~\ref{fig:lifetime-boxplot}, where we readily notice that the hardware implementation of AES-128 significantly outperforms the software AES-128 implementation, plus other \textit{lightweight} software ciphers such as SPECK and MISTY1. It is also interesting to observe that the energy impact of the hardware AES-128-based protocol is not significant when comparing with an unsecured communication.


\Figure[!t]()[trim={0cm 0cm 0cm 0.8cm},clip,scale=0.55]{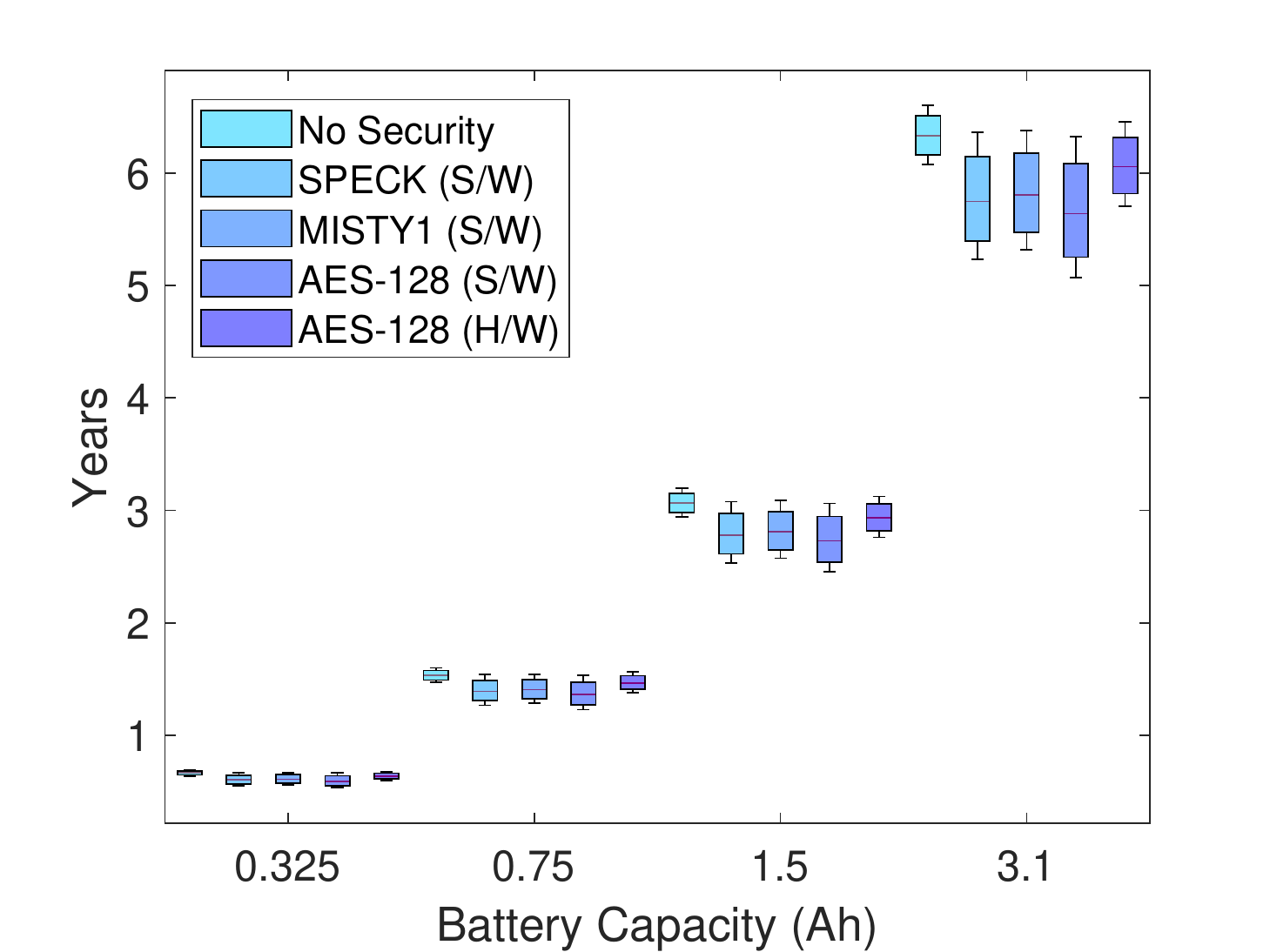}
{IMD-battery lifetime with respect to cryptographic primitive used. Boxplot variation is due to different data-transfer volumes\label{fig:lifetime-boxplot}}

\subsection{IMD performance}
\label{sec:performance}

To study the impact of \simdprotocol on performance during normal operation, we will only analyze the bottleneck of the reader-IMD system in this regard, i.e., the IMD itself.
This is because modern readers, such as tablets~\cite{proclaim2017proclaim}, have far superior computational resources (and battery autonomy) than implants. As far as the smart card is concerned, the amount of computations performed by it is approximately the same as that in commercial uses (e.g., EMV), which we know to exhibit adequate performance.

As far as the IMD is concerned, the performance figure of merit that is crucial to capture here is the delay that \simdprotocol incurs to the system, both for security computations and data transmission over the air. For unsecured data transfer, the wireless transceiver incurs a delay of 2.2 ms. As shown in Fig.~\ref{fig:plot_emv_imd}, for (hardware-accelerated) secure data transfer the time delay incurred by each (numbered) protocol step is no higher than 6 ms, for a total protocol delay of 15.7 ms. Therefore, for the time scales involved in biological processes, we can safely assume that the \simdprotocol delay overhead is negligible.

\subsection{Summary of introduced overheads}
Table~\ref{table:final_summary} summarizes the impact of \simdprotocol on an IMD in terms of energy, performance and program-memory footprint.
For the hardware implementation of \simdprotocol, it can be observed that, although the energy requirements increase by more than 6 times for a basic session, the \textit{total} daily IMD consumption (that includes a two-minute communication session and electrical-stimulation costs) increases from 16.60 J to just 17.69 J, which amounts to a mere 6.57\% increase, as previously shown in Fig.~\ref{fig:lifetime-boxplot}.
The reason for this small increase is that the basic medical functionality, e.g., the continuous electrical stimulation of a pacemaker, dominates the security provisions since the reader accesses are far less frequent.
In the case of software (AES-128) implementation of \simdprotocol, the total daily IMD consumption increases by 19.82\% (as shown in Fig.~\ref{fig:lifetime-boxplot}).
Moreover, there is a minimal increase in the computational delay and required program-memory size.
In the context of current MCU technology, 8.22--10.48 kB of additional memory size is negligible.
Hence, we conclude that there is no noticeable change in the IMD costs when \simdprotocol is employed.

\begin{table}
	\begin{threeparttable}
    \caption{Summary of costs for running the \simdprotocol protocol on an IMD}
    \label{table:final_summary}
    \begin{tabular}{@{}lrrrr@{}}
        \toprule
        & \multicolumn{2}{c}{Energy}            & Delay & Prog-Mem \\\cline{2-3}
        & 1 basic					& 1 daily IMD &       & footprint$^{**}$      \\
        & session ($\mu$J)        & cycle$^*$~(J)          & (ms)  & (kB)         \\
        \midrule
        Without security   & 16.61         & 16.60             & 2.17   & 16.50   \\
        \simdprotocol (H/W) & 108.31        & 17.69            & 15.73  & 24.72   \\
        \simdprotocol (S/W) & 217.89        & 19.89            & 58.99  & 26.98   \\
        \bottomrule
    \end{tabular}
	\begin{tablenotes}
		\footnotesize
		\item [*] Which includes a daily two-minute comm. session (see Section~\ref{sec:availability})
		\item [**] This includes the comm. data handling, security processing and MCU peripheral support
		library for GPIO and USART, which are needed to communicate with the transceiver.
	\end{tablenotes}
\end{threeparttable}
\end{table}

\section{Conclusions}
\label{sec:conclusion}

In this paper, we have proposed a novel security protocol for IMD ecosystems, \simdprotocol.
We have demonstrated that our approach offers a meticulous coverage of security requirements that are critical to these systems.
This becomes possible through the use of a personal smart card and a trusted third party, which helps in facilitating access control, non-repudiation, user authentication, bedside-reader operation and system scalability.
We have also shown that \simdprotocol does not introduce any noticeable overheads in the implant, and it has the ability to support zero-power defense against battery-DoS attacks.
It is observed that our proposed protocol increases the total IMD energy consumption by just 6.57\%, which is minimal in the context of the IMD lifespan.
We have also proposed an OOB-channel-based version of \simdprotocol, which enables offline or emergency access.

\bibliographystyle{IEEEtran}
\bibliography{ms}

\begin{IEEEbiography}[{\includegraphics[width=1in,height=1.25in,clip,keepaspectratio]{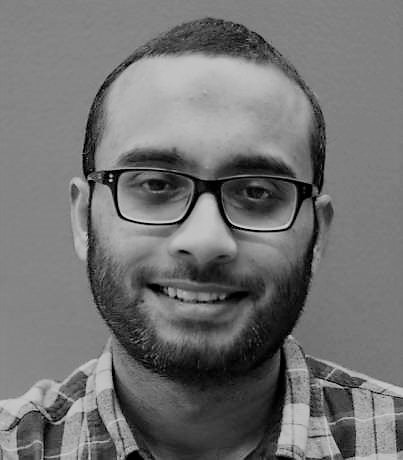}}]{Muhammad Ali Siddiqi} received the B.E. degree in electrical (telecommunication) engineering from National University of Sciences and Technology, Islamabad, Pakistan, in 2009 and the (joint) M.Sc. degree in embedded computing systems from Norwegian University of Science and Technology, Trondheim, Norway, and University of Southampton, UK, in 2012. He is currently pursuing the Ph.D. degree at the Neuroscience department of the Erasmus Medical Center, the Netherlands.

From 2012 to 2017, he worked as a Design Engineer at Silicon Labs Norway on the ultra-low-power MCU design. His research interest includes the development of security protocols and architectures for heavily resource-constrained embedded systems, such as implantable medical devices.
	
\end{IEEEbiography}

\begin{IEEEbiography}[{\includegraphics[width=1in,height=1.25in,clip,keepaspectratio]{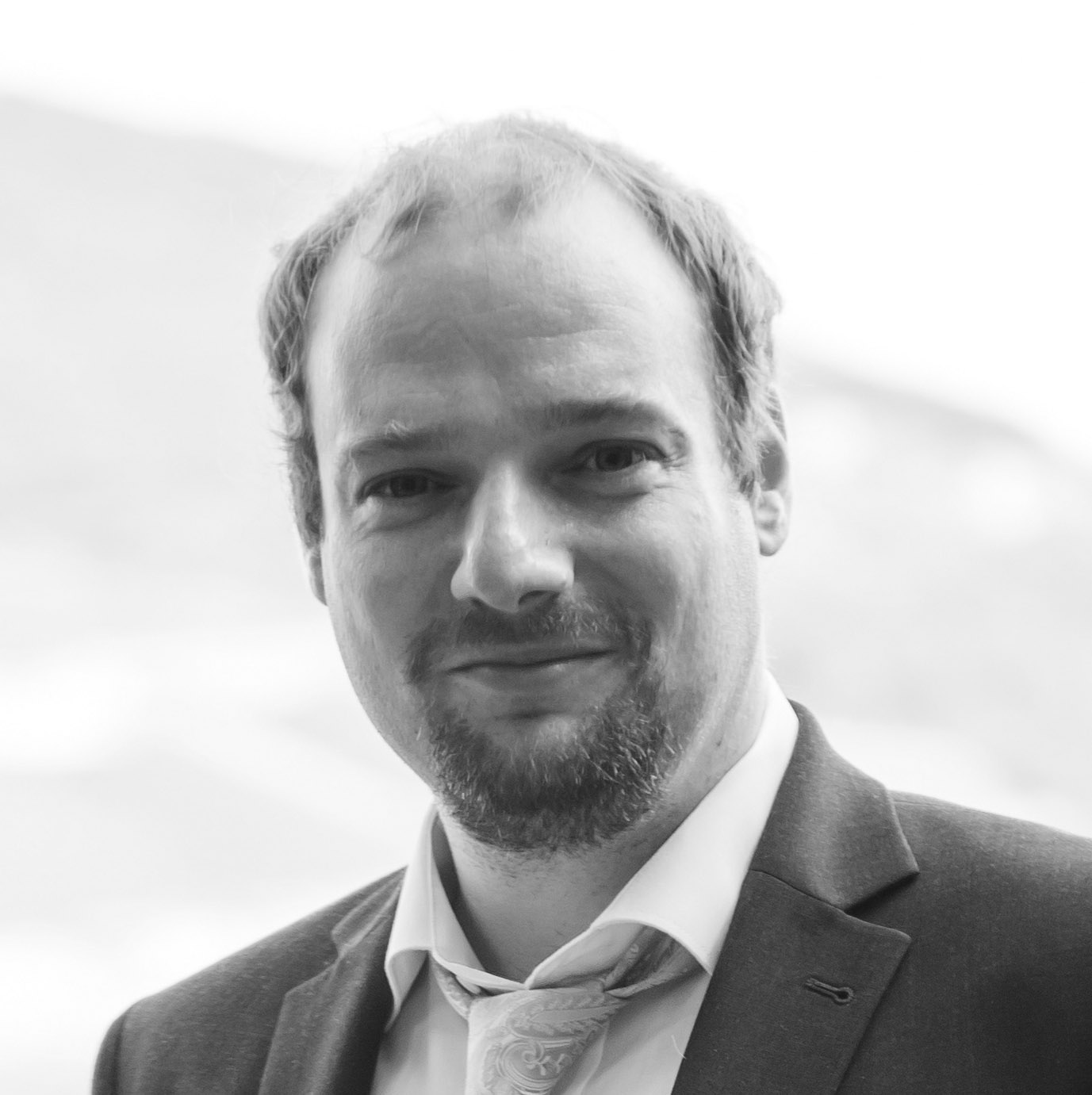}}]{Christian Doerr} is a professor at the Hasso Plattner Institute, University of Potsdam, Germany. He specializes in network security, specifically cyber threat intelligence and situational awareness, as well as protection of critical information infrastructures. He has received his Ph.D. in Computer Science and Cognitive Science from the University of Colorado at Boulder in 2008. Prof. Doerr is PI of the Cyber Threat Intelligence Lab, which analyses techniques and tactics used by adversaries. His group also operates a network telescope, which is used to track and quantify the nature and type of attacks on the Internet. He served as the TPC chair of the International Conference on Availability, Reliability and Security (ARES 2018) and is the initiator of WCTI, the first International Workshop on Cyber Threat Intelligence, an event specifically targeted towards exchange of threat information and defense techniques.
\end{IEEEbiography}

\begin{IEEEbiography}[{\includegraphics[width=1in,height=1.25in,clip,keepaspectratio]{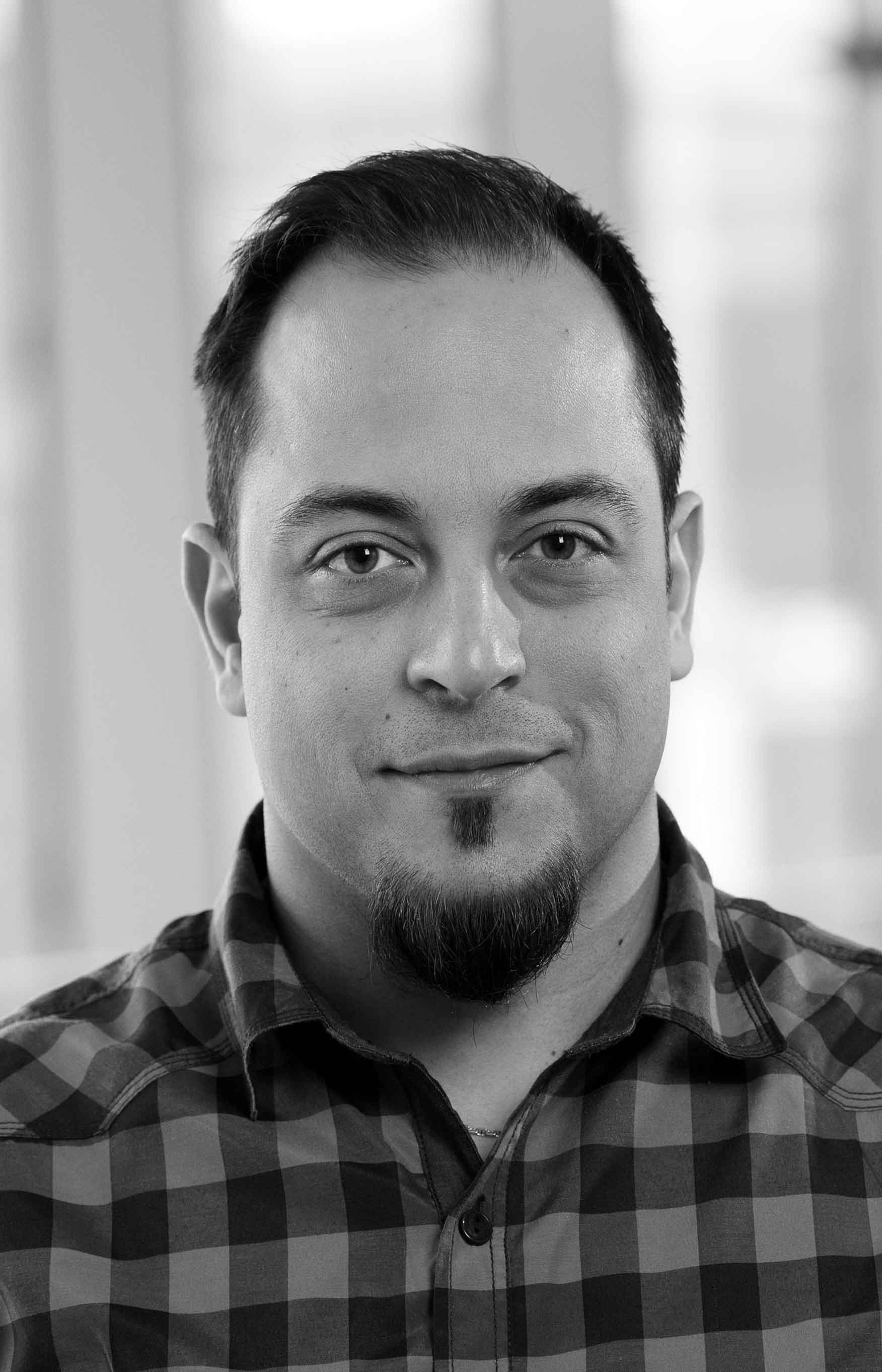}}]{Christos Strydis} (Senior Member, IEEE) received the M.Sc. \textit{(magna cum laude)} and the Ph.D. degrees in computer engineering from the Delft University of Technology. He is currently a tenured Assistant Professor in computer engineering and the Head of the NeuroComputing Laboratory, Neuroscience Department, Erasmus Medical Center, The Netherlands. He has published work in well-known international conferences and journals. He has delivered invited talks in various venues. His current research interests include brain simulations, high-performance computing, low-power embedded (implantable) systems, and functional ultrasound imaging.
	
\end{IEEEbiography}

\EOD

\end{document}